\documentclass[12pt,preprint]{aastex}
\usepackage{rotating}
\usepackage{graphicx}

\shorttitle{The SDSS/XMM-Newton Quasar Survey}
\shortauthors{M. Young et al.}

\begin{document}

\title{The X-ray Energy Dependence of the Relation between Optical and X-ray Emission in Quasars}

\author{
M. Young\altaffilmark{1,2}, M. Elvis\altaffilmark{1}, 
G. Risaliti\altaffilmark{1,3}
} 
\email{myoung@cfa.harvard.edu}

\altaffiltext{1}{Harvard-Smithsonian Center for Astrophysics, 60 Garden St. 
Cambridge, MA 02138 USA}
\altaffiltext{2}{Boston University, Astronomy Department, 725 Commonwealth Ave., 
Boston, MA 02215}
\altaffiltext{3}{INAF - Osservatorio di Arcetri, L.go E. Fermi 5,
Firenze, Italy}
\begin{abstract}
We develop a new approach to the well-studied anti-correlation between the optical-to-X-ray spectral index, 
$\alpha_{ox}$, and the monochromatic optical luminosity, $l_{opt}$.  By cross-correlating the SDSS DR5 quasar 
catalog with the XMM-Newton archive, we create a sample of 327 quasars with X-ray S/N $>$ 6, where both optical 
and X-ray spectra are available.  
This allows $\alpha_{ox}$ to be defined at arbitrary frequencies, rather than the standard 2500 $\mbox{\AA}$ 
and 2 keV.  We find that while the choice of optical wavelength does not strongly influence the $\alpha_{ox}-
l_{opt}$ relation, the slope of the relation does depend on the choice of X-ray energy.  The slope of the 
relation becomes steeper when $\alpha_{ox}$ is defined at low ($\sim$ 1 keV) X-ray energies.  This change 
is significant when compared to the slope predicted by a decrease in the baseline over which $\alpha_{ox}$ is 
defined.  The slopes are also marginally flatter than predicted at high ($\sim$ 10 keV) X-ray energies.  
Partial correlation tests show that while the primary driver of $\alpha_{ox}$ is $l_{opt}$, the Eddington 
ratio correlates strongly with $\alpha_{ox}$ when $l_{opt}$ is taken into account, so accretion rate may 
help explain these results.
We combine the $\alpha_{ox}-l_{opt}$ and $\Gamma$-L$_{bol}$/L$_{Edd}$ relations to naturally 
explain two results: 1) the existence of the $\Gamma-l_x$ relation as reported in \citet{Young09} and 2) 
the lack of a $\Gamma-l_{opt}$ relation.  The consistency of the optical/X-ray correlations establishes 
a more complete framework for understanding the relation between quasar emission mechanisms.  
We also discuss two correlations with the hard X-ray bolometric correction, which we show correlates with 
both $\alpha_{ox}$ and Eddington ratio.  This confirms that an increase in accretion rate correlates with 
a decrease in the fraction of up-scattered disk photons.
\end{abstract}

\keywords{ Galaxies: AGN --- accretion disks --- X-rays: general}

\section{Introduction}

Optical/UV and X-ray continua in quasars are thought to originate in two physically distinct, 
but nevertheless related regions.  The optical/UV spectrum is dominated by a ‘Big Blue Bump’ 
continuum feature \citep[e.g., ][]{Elvis94}, most likely from thermal emission arising from 
an accretion disk \citep{Shields78, MS82, Ward87}.  A hot gas lying above the disk in some 
unknown geometry then up-scatters these disk photons to form a hard power-law ($<\alpha_x> 
\sim$ -1) \citep[e.g.,][]{Mateos05, Mainieri07} in the X-ray spectrum \citep{HM91, Zdziarski00, 
Kawaguchi01}.  

While the shapes of the optical and X-ray continua are well known, the connecting continuum in 
the extreme ultraviolet (EUV) is not, because of absorption from our own galaxy.  The EUV and 
soft X-rays form the ionizing continuum (h$\nu >$ 13.6 eV) needed to ionize a quasar's 
characteristic broad lines.  The ionizing continuum is also responsible for any line-driven 
outflows from the accretion disk \citep{Proga07}.  Since direct observation of this continuum 
is not possible, it is often parameterized by an imaginary power-law with spectral index 
$\alpha_{ox}$ running from 2500 $\mbox{\AA}$ to 2 keV in the rest frame.  

The importance of the ionizing continuum motivates the decades long study of the anti-correlation between 
$\alpha_{ox}$ and L$_{2500}$ \citep{AT82, KC85, Tananbaum86, AM87, Wilkes94, PIF94, AWM95}; 
however, controversy continues.  While some studies have found that $\alpha_{ox}$ depends primarily 
on redshift rather than optical luminosity \citep{Bechtold03, Kelly07}, others have suggested that 
$\alpha_{ox}$ may be entirely independent of redshift and optical luminosity \citep{Yuan98, Tang07}, 
arguing that if the dispersion in L$_{2500}$ is much greater than the dispersion in L$_{2keV}$, an 
apparent but artificial correlation will result.  A narrow range in optical luminosity will enhance 
this effect.  

In recent years, studies of the $\alpha_{ox}-L_{2500}$ relation have become more comprehensive, 
with higher X-ray detection rates and largely complete samples that span a wide range of luminosities 
\citep{Vignali03, Strateva05, Steffen06, Just07}.  For example, \citet{Steffen06} work spans five and four 
decades in UV and X-ray luminosity, respectively, which helps minimize the effect of an artificial 
$\alpha_{ox}-L_{2500}$ correlation due to selection effects.  In addition, \citet{Strateva05} estimated 
the dispersion in L$_{2500}$ and L$_{2keV}$, finding that the difference in dispersions was not enough 
to artificially reproduce a correlation.  These studies used partial correlation analysis to conclude 
that the $\alpha_{ox}-$z correlation is insignificant, while the $\alpha_{ox}-L_{2500}$ anti-correlation 
is highly significant.  However, 
while the significance of the relation continues to be better understood, little is known of the 
underlying physics.  Current disk-corona models \citep{Beloborodov99, Zdziarski99, Nayakshin00, Malzac01, 
Sobolewska04a, Sobolewska04b} do not directly predict the $\alpha_{ox}-L_{2500}$ relation.  

Previous studies of the $\alpha_{ox}-L_{2500}$ relation have largely lacked spectral slopes in the X-rays, 
and so used the traditional endpoints of 2500~$\mbox{\AA}$ and 2 keV, where an X-ray spectral slope ($\Gamma 
\sim$ 2, where $\Gamma$ = -$\alpha$ + 1 for F$_{\nu} \propto \nu^{\alpha}$) is assumed in order to obtain 
the X-ray flux at 2 keV.  A systematic study with both optical and X-ray spectra enables an investigation 
of the relation at different frequencies than those traditionally used, revealing clues about the relation's 
physical underpinnings.  

The sample selection and measurement of $\alpha_{ox}$ are described in \S2, and results of the data analysis 
are given in \S3.  Section 4 discusses the results, specifically their connection to other optical/X-ray 
relations found for quasars.  We assume a standard cosmology throughout the paper, where H$_0$ = 70 km 
s$^{-1}$ Mpc$^{-1}$, $\Omega_M = 0.3$, and $\Omega_\Lambda = 0.7$ \citep{Spergel03}. 

\section{The Sample}

We have cross-correlated the DR5 Sloan Digital Sky Survey (SDSS) with \emph{XMM-Newton} archival 
observations to obtain 792 X-ray observations of SDSS quasars.  This gives the largest sample of 
optically selected quasars with both optical and X-ray spectra (473 quasars with X-ray S/N $>$ 6).  
\citet{Young09} describes the SDSS quasar selection and cross-correlation with the \emph{XMM-Newton} 
archive, as well as the X-ray data reduction and spectral fitting.  For the purposes of this paper, 
we exclude broad absorption line (BAL) and radio-loud (RL) quasars, as these are known to have 
different $\alpha_{ox}$ distributions than typical, radio-quiet (RQ) quasars \citep{GM96,Young09}.  
While \citet{Young09} used an incomplete BAL list \citep{Shen08}, this paper updates the BAL 
classification using the DR5 BAL catalog \citep{Gibson09}.  A total of 18 sources are re-classified 
as BALs, and an additional 15 sources are re-classified as non-BALs.  A net result of three additional 
BALs, gives the SDSS/XMM-Newton Quasar Survey 55 BALs in total.  

Since the goal is to use our knowledge of the X-ray spectrum to determine the effects of the X-ray 
continuum shape on the calculation of $\alpha_{ox}$, we form a sample of quasars with enough 
X-ray S/N to fit a spectrum (S/N $>$ 6) in the primary sample of this study, hereafter referred 
to as SPECTRA.  As in \citet{Young09}, we fit three models to each spectrum: 1) a single power-law 
(SPL) with no intrinsic absorption, 2) a fixed power-law (FPL) with intrinsic absorption left free 
to vary, and 3) an intrinsically absorbed power-law (APL), with both photon index $\Gamma$ and 
absorption N$_H$ left free to vary.  Any spectrum that does not give a good fit with any of the 
above models is excluded from the sample, in order to ensure well-determined X-ray fluxes.  Therefore 
the final sample will not include any sources with, for example, a significant contribution from a 
strong soft excess component.  In addition, any spectrum that prefers either of the absorbed models 
(FPL or APL) is excluded from the sample, in order to minimize the effect of absorption on both X-ray 
and optical flux  measurements.  This reduces the sample by 55 sources, or 11.6\% of quasars with X-ray 
S/N $>$ 6.  Once all of the selections are taken into account, the sample contains 327 quasars 
spanning a redshift range of z = 0.1 - 4.4.  

The median upper-limit for intrinsic absorption in the SPECTRA sample is N$_H \lesssim$ 4 x 10$^{21}$ 
cm$^{-2}$.  To check for the effect of undetected absorption in low S/N X-ray spectra, we raise the 
threshold to S/N $>$ 20, where N$_H$ upper limits decrease to $\sim$ 6 x $10^{20}$ cm$^{-2}$.  This 
sample contains 99 sources and will hereafter be referred to as HIGHSNR.

Finally, we avoid bias against X-ray weak sources by adding undetected or low-signal sources to the 
SPECTRA sample, to create the CENSORED sample.  To include sources with S/N $<$ 6, we apply a fixed 
power-law fit with no absorption, where the power-law is fixed to $\Gamma$ = 1.9, the average value 
for the SDSS/XMM-Newton Quasar Survey \citep{Young09}.  RL and BAL quasars are still excluded, as are 
any sources with bad X-ray fits, leaving 514 sources in the sample spanning a redshift range of z = 0.1 - 5.0.  

\begin{figure}
\centering
\includegraphics[]{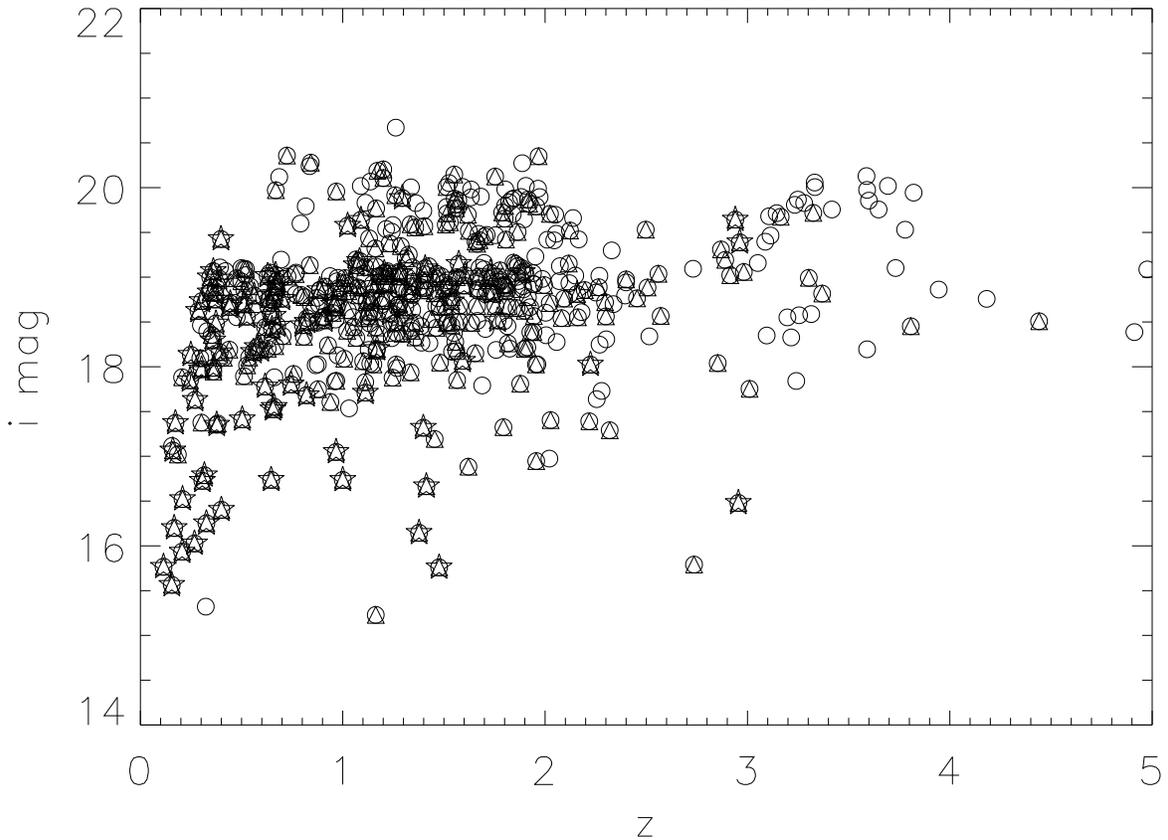}
\caption{The i band magnitude vs. redshift for three samples: CENSORED (circles), 
SPECTRA (triangles), and HIGHSNR (stars).}
\label{fig:imagvsz}
\end{figure}

The three samples used in this paper are described in Table 1.  Figure \ref{fig:imagvsz} 
gives the $i$ band magnitude vs. redshift distribution of each of the three samples.

\subsection{Measuring $\alpha_{ox}$}

For each SPECTRA sample source, we fit six versions of an unabsorbed power-law.  The normalization 
is frozen to rest-frame energy values of 1, 1.5, 2, 4, 7 and 10 keV and then the monochromatic 
fluxes are taken at each energy.  Freezing the normalization at each energy enables us to 
accurately calculate the error on the flux based on the error on the normalization of the fit.  
If the normalization was frozen to 1 keV observed-frame, as is standard, both the error in 
normalization and the error on $\Gamma$ would affect the error on the flux.  For 215 quasars, 
at least one of the rest-frame energies lies below the observed range.  In these cases, 
we normalize the spectrum to 1 keV and extrapolate to obtain the flux at the rest-frame energy.

We do not normalize to rest-frame energies lower than 1 keV because undetected absorption with 
a median upper limit of N$_H \sim$ 5 x 10$^{21}$ would eliminate any measured flux at 0.5 keV, 
and would reduce the flux measured at 0.75 keV by an order of magnitude.  At 1 keV, the change 
in flux is only a factor of $\sim$2, which would change $\alpha_{ox}$ by $\sim$0.1.  Limiting 
the normalization energies to 1 keV and higher has the added benefit of minimizing the contribution 
from any soft excess component.  

We also do not attempt to fit fluxes at rest-frame energies higher than 10 keV, as these are only 
directly available for high redshift (and therefore high luminosity) sources.  Extrapolation with 
the fitted power-law for low redshift sources would not take into account the reflection bump that 
begins rising at 10 keV and peaks near 30-50 keV.  For low redshift sources, rest-frame energies of 
7-10 keV will sample the 7-10 keV observed spectrum, where under- or over-subtraction of the background 
can alter the measured fluxes.  The resulting increase in flux measurement error will affect the 
dispersion of the $\alpha_{ox}-l{opt}$ relation, which will be discussed further in \S3.  

Before calculating optical luminosities, we first correct the magnitudes for Galactic reddening.  
Corrections are applied using a Galactic dust-to-gas ratio and the Galactic hydrogen column density 
from WebPIMMS\footnote{http://heasarc.gsfc.nasa.gov/Tools/w3pimms.html}, which is based on the 21 cm 
HI compilation of Dickey \& Lockman (1990) and Kalberla et al. (2005)\nocite{DL90,Kalberla05}.  We 
then interpolate a power-law through the well-calibrated SDSS photometry ($\sim$2-3\% photometric error).  
As detailed below, we carefully select wavebands for interpolation to avoid host galaxy contribution 
at low redshift, the Ly$\alpha$ forest at high redshift, and major emission lines.  

For redshifts z $<$ 0.55, the host galaxy can redden the magnitudes at long wavelengths, so we use the 
bluest bands, $u$, $g$ and $r$, to determine the optical slope, while also avoiding those bands containing 
the MgII line.  For z $<$ 0.25, we use the $g$ and $r$, switching to $u$ and $g$ for 0.25 $<$ z $<$ 0.43, 
and then to $u$ and $r$ for 0.43 $<$ z $<$ 0.55.  (Note that the MgII line lies largely between the $u$ 
and $g$ filters for the middle redshift range, 0.25 $<$ z $<$ 0.43.)  

For quasars at redshifts above z = 0.55, we assume host galaxy contribution to be minimal 
\citep{Richards03} and fit a power-law through all wavebands using an ordinary least-squares regression.  
While lines such as MgII, CIII and CIV affect the photometry at various redshifts, their addition to the 
continuum flux is minimized by the broad bandwidths (the $g$,$r$,$i$ and $z$ bands have FWHM $\sim$ 1000 
$\mbox{\AA}$).  In addition, the regression fit to all wavebands minimizes the contribution of any one 
waveband that is an outlier due to line emission.  

As the Ly$\alpha$ line begins to affect the wavebands, they are removed from the regression 
fit.  The $u$ band is removed from the fit at z = 1.8, and the $g$ band at z = 2.3.  At z $>$ 3.75, for 
two quasars, the Ly$\alpha$ line affects the spectrum significantly and we simply fix the optical slope to 
the median value from \citet{VdB01}, $\alpha_{opt}$ = 0.46.  

The normalization of the optical power-law is also adjusted according to redshift.  For sources with z $<$ 
0.55, the power-law is normalized at the bluest band to avoid host galaxy contamination.  For sources with 
0.55 $<$ z $<$ 3.75, the power-law is normalized using the ordinary least squares fit to all of the available 
magnitudes.  For the single source with z $>$ 3.75, the power-law is normalized to the $z$ band to minimize 
the effect of the Ly$\alpha$ forest.  From the measurement of the index and normalization of the optical 
power-law, we can extrapolate to three monochromatic optical luminosities at 1500, 2500 and 5000 $\mbox{\AA}$, 
where the log of the optical luminosity is hereafter denoted as $l_{opt}$.  

The blend of FeII and Balmer emission lines that make up the Small Blue Bump \citep{WNW85} affects 
monochromatic optical fluxes by 0-45\% \citep{BG92}.  This causes at most a small overestimate of 
$\alpha_{ox}$ by $\Delta$($\alpha_{ox}$) $\sim$ 0.06.  

The mean optical slope obtained for the SDSS/XMM-Newton Quasar Survey ($\alpha_{opt}$ = -0.40) is similar 
to the slope of the \citet{VdB01} mean composite ($\alpha_{opt}$ = -0.44), indicating that the host galaxy 
and emission line contributions have been adequately taken into account.  The dispersion of the optical 
slope is $\sigma_\alpha$ = 0.30, similar to the dispersion seen in the composites of \citet{Richards03}, 
where $\sigma_\alpha$ $\sim$ 0.26.  

Once the optical luminosities are obtained, corrections for dust-reddening are applied where necessary.  
We obtain an estimate of E(B-V) from the relative ($g - i$) color, where relative colors compare a quasar's 
measured colors with the median colors in its redshift bin, so that $\Delta$($g - i$) = ($g - i$)~-~$<$($g - 
i$)$>_z$ \citep{Richards03}.  Quasars with $\Delta$($g - i$) $<$ 0.2 lie within the normal color distribution 
for SDSS quasars, and so are not likely to be affected by dust-reddening.  For quasars with $\Delta$($g - i$) 
$>$ 0.2 the E(B-V) estimated from the relative colors is used to de-redden the optical luminosities with 
an SMC-type extinction curve \citep{Prevot84}.  Note that the correction can only be made for sources with 
redshift less than 2.2 for the relative colors to be valid, due to the entry of the Ly$\alpha$ line into the 
spectrum.  Sources with E(B-V) $>$ 0.04 are classified as dust-reddened and make up 7.6\% of the SDSS/XMM-Newton 
Quasar Survey.  This fraction is comparable to that found for SDSS quasars \citep[6\%, ][]{Richards03}.  

We next calculate $\alpha_{ox}$ with the standard formula: 
\begin{displaymath} \alpha_{ox} = \frac{log(f_x/f_o)}{log(\nu_x/\nu_o)}\end{displaymath} 
where $f_x$ is the monochromatic flux at 1, 1.5, 2, 4, 7, or 10 keV and $f_o$ is the 
monochromatic flux at 1500, 2500, or 5000 $\mbox{\AA}$.

For each $\alpha_{ox}-l_{opt}$ relation, we fit an ordinary least-squares regression line 
and obtain the slope and intercept.  The errors are at the 1$\sigma$ level, and the standard 
deviation on the regression gives the observed scatter in the relation.  

For the CENSORED sample, we fix the power-law to the sample average of $\Gamma$ = 1.9 for 187 
sources with S/N $<$ 6, and then fit the monochromatic X-ray flux.  For 59 undetected sources 
with S/N $<$ 2, we fit the 90\% upper-limit on the flux.  Since the data now include upper 
limits, we input the lists of $\alpha_{ox}$ and $l_{opt}$ into the `Estimate and Maximize' 
(EM) function to compute the linear regression.  This method is among several routines available 
on ASURV  Rev 1.2, a survival statistics package \citep{Lavalley92}, which implements the methods 
presented in \citet{Isobe86}.  Survival statistics account for censored data by assuming that 
the upper limits follow the same distribution as the detected points.  The EM regression reduces 
to an ordinary least squares fit for uncensored data.  

For all three samples, we use a partial correlation test designed to measure the correlation 
between two variables, while controlling for the effects of a third.  The FORTRAN program 
CENS\_TAU calculates Kendall's $\tau_{12,3}$ partial correlation coefficient, and can take 
censored data into account using the methodology presented in \citep{AS96}.  Unless otherwise 
noted, redshift is the third variable for the correlation tests.  Both ASURV and 
CENS\_TAU are made available by the Penn State Center for 
Astrostatistics\footnote{http://www.astrostatistics.psu.edu/statcodes/cens\_tau}.  
 
\section{Results}

We first measure the $\alpha_{ox}-l_{opt}$ anti-correlation using the analysis of previous 
studies \citep{Vignali03, Strateva05, Steffen06} on the CENSORED sample.  In comparison to 
\citet{Steffen06}, the CENSORED sample is larger (N=514, compared to N=333), with a similar 
range of $\sim$4 decades in X-ray luminosity.  The X-ray upper limit fraction is similar as well,
(11\%, compared to 12\% for \citet{Steffen06}).  The range in optical luminosity in \citet{Steffen06}
is roughly 1.5 decades larger than the range in this paper (3.5 decades), due largely to 
their inclusion of low and moderate-luminosity samples in addition to the SDSS.  

\begin{figure}
\centering
\includegraphics[]{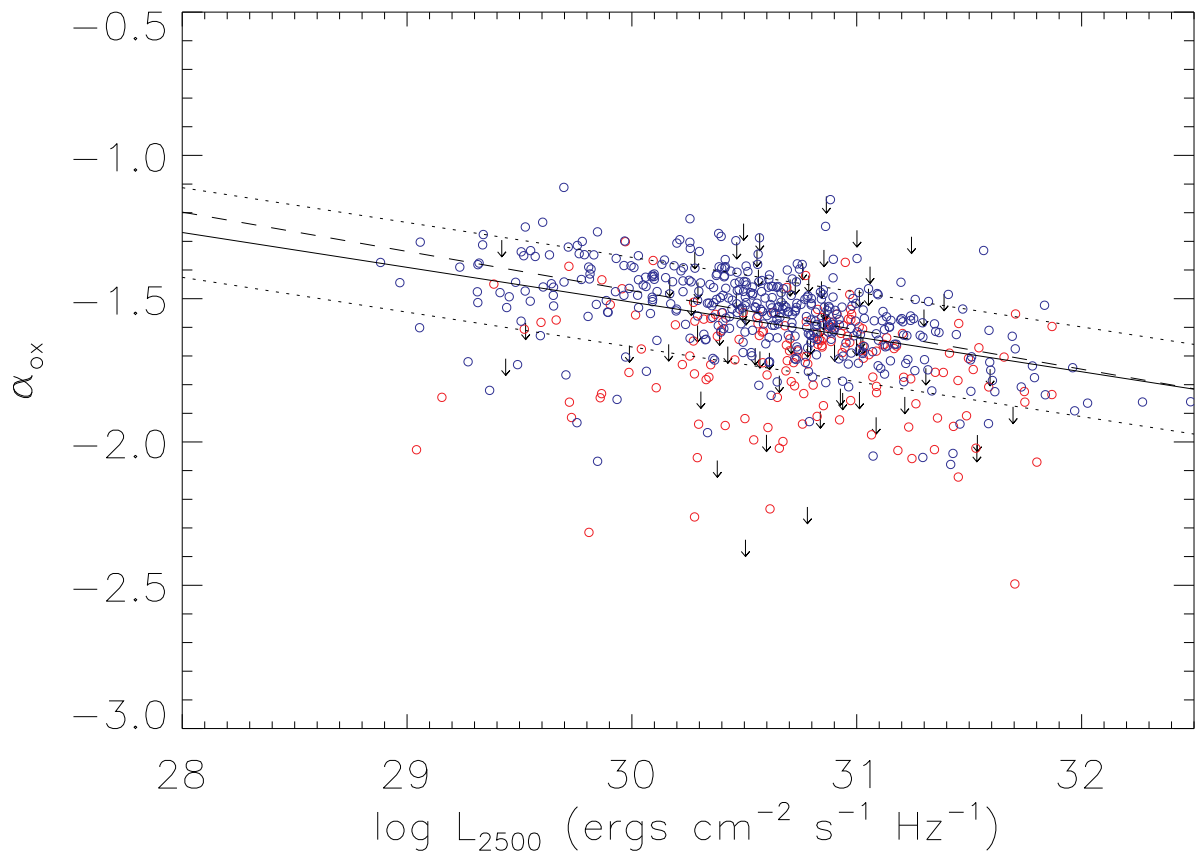}
\caption{The $\alpha_{ox}-l_{opt}$ relation for the CENSORED sample.  Blue circles indicate sources with 
S/N $>$6 (with a power-law fit to the X-ray spectrum), red circles indicate detected sources with S/N $<$ 6 
(no spectral fit, only X-ray fluxes), and black arrows indicate undetected sources with X-ray upper-limits.  
The solid line is a linear least-squares fit to the data, with dispersions plotted as dotted lines.  The 
dashed line shows the \citep{Steffen06} best-fit line.}
\label{fig:aoxvsLopt_old}
\end{figure}

Figure \ref{fig:aoxvsLopt_old} shows the $\alpha_{ox}-l_{opt}$ anti-correlation, significant 
at the 9.0$\sigma$ level ($\tau_{12,3}$ = -0.20) in the \emph{SDSS/XMM} Quasar Survey 
when defined at the conventional 2500 $\mbox{\AA}$ and 2 keV fiducial points.  The solid line 
is the best-fit EM regression line for the CENSORED sample: 
\begin{displaymath} \alpha_{ox} = (2.133 \pm 0.388) - (0.122 \pm 0.013) log(L_{2500\AA}) 
\end{displaymath}  
The slope is consistent within errors to the best-fit regression from \citet{Steffen06} 
(dashed line in Fig. \ref{fig:aoxvsLopt_old}).  Unlike in \citet{Steffen06}, we are not 
able to reject a correlation between $\alpha_{ox}$ and redshift while controlling for 
$L_{2500\AA}$.  Instead we find the partial correlation to be marginally significant at 
the 3.8$\sigma$ level ($\tau_{12,3}$ = -0.085) in the CENSORED sample.  This is likely 
due to the smaller range in optical luminosity, as well as the large scatter in $\alpha_{ox}$.  
Note that in the SPECTRA sample, the partial correlation between $\alpha_{ox}$ and redshift 
is largely insignificant, except at 1 keV where, again, the scatter is significant (see Table 
\ref{tab:aoxz}).  

We can now test the correlations of $\alpha_{ox}$ versus optical luminosity using the 
SPECTRA sample, for each of the six X-ray energies and three optical wavelengths where 
$\alpha_{ox}$ is defined.  Figure 3 shows 6 examples of the relation, defined at 1500, 
2500 and 5000 $\mbox{\AA}$ and at 1 and 10 keV.  The plots show a trend in the slope and 
dispersion of $\alpha_{ox}-l_{opt}$ relation: the slope flattens and the dispersion tightens 
as the X-ray energy used to define $\alpha_{ox}$ increases.  It is clear from the plots that 
the effect of changing X-ray energy is much stronger than that of changing optical wavelength.  

The definition of $\alpha_{ox}$ depends on the baseline over which it is defined; lengthening 
the baseline can produce an artificial flattening of the $\alpha_{ox}-l_{opt}$ slope.  To check 
that the intrinsic slope of the correlation is changing, we define the standard slope by the 
relation at 2500 $\mbox{\AA}$ and 2 keV.  We then calculate how the change in baseline affects 
the slope and intercept of the relation.  As the changing baseline frequencies are accompanied 
by a change in the X-ray flux, we include this effect by assuming an average value for the X-ray 
slope ($<\alpha_x> = -0.9$).  The equations for the change in slope (m) and intercept (b) then  become:
\begin{displaymath} m' = m\:\frac{log(\nu_{2keV}/\nu_{2500\AA})}{log(\nu_x/\nu_o)} 
\end{displaymath} 
and
\begin{displaymath} b' = \frac{1}{log(\nu_x/\nu_o)} \left[\, b \, log(\nu_{2keV}/\nu_{2500\AA}) 
+ <\!\alpha_x\!> log(\nu_x/\nu_{2keV}) \right]\end{displaymath}  
where $\nu_x$ is the X-ray frequency at 1, 1.5, 2, 4, 7, or 10 keV and $\nu_o$ is the 
optical frequency at 1500, 2500, or 5000 $\mbox{\AA}$.

A change in the optical baseline, for example from 2500 to 5000 $\mbox{\AA}$, affects the 
slope of the $\alpha_{ox}-l_{opt}$ relation by an amount comparable to the error.  This is 
in contrast to \citet{Vasudevan09}, who find a marginal strengthening of the $\alpha_{ox}-
l_{opt}$ relation as the optical reference point is moved to shorter wavelengths.  This may 
be due to a steepening of the correlation slope, but the small sample size and the strong 
effect of reddening on the sample do not allow a definitive statement.  
Since the slope does not depend significantly on optical wavelength in our results, we plot 
our results only for 2500 $\mbox{\AA}$ in Figure 4, in order to focus on the change with 
X-ray energy.  

Figure 4$a$ shows that while changing the X-ray baseline clearly has an influence on 
the measured slopes, the slope for 1 keV is significantly steeper than it would be 
if the intrinsic correlation were constant.  The slopes at 4, 7 and 10 keV are flatter 
than predicted by the baseline effect at the 1$\sigma$ level.  However, since the 
baseline effect is arbitrarily normalized to the slope at 2 keV, the main result of 
subtracting the baseline effect is not to assess the significance of the slope at a 
given energy, but rather to show that the $intrinsic$ slope of the $\alpha_{ox}-l_{opt}$ 
relation is flattening as X-ray energy increases.

Figure 4$b$ and 4$c$ show the intercept and dispersion of each relation as measured via 
the least squares fit.  The intercept of the relation decreases significantly with X-ray 
energy compared to the predicted curve.  The dispersion of the $\alpha_{ox}-l_{opt}$ 
relation largely follows the predicted curve, though it is larger than predicted at the 
lowest and highest energies.  This is likely due to the larger errors in X-ray flux at 
the ends of the spectrum.  

For low redshift sources, rest-frame energies of 7-10 keV will sample the 7-10 keV observed 
spectrum, where under- or over-subtraction of the background can alter the measured fluxes 
and increase the error in the flux measurement.  When we exclude sources with z $<$ 0.5, we 
find that the dispersion of the $\alpha_{ox}-l_{opt}$ relation at 7 and 10 keV decreases by 
$\sim$ 0.005, though the dispersions do not decrease all the way to the baseline level.  
This suggests that the background does have some effect at low redshifts, but it is not the 
only source of scatter at high energies.

Interestingly, the trend in slope and dispersion of the $\alpha_{ox}-l_{opt}$ relation for 
the CENSORED sample is mostly consistent with the baseline effect (Fig. \ref{fig:CENSORED}), 
with a significantly steeper slope than expected at 1 keV.  The slopes also show the 
same marginal flattening at high X-ray energies as the SPECTRA sample.  These results are 
not as strong as in the SPECTRA sample.  

Figure \ref{fig:aoxvsLopt_old} suggests that the main effect of including low S/N and 
undetected sources is an increase in scatter, diluting any potential trends.  The red circles 
in the figure, representing detected sources with no spectral fit, show a weaker anti-correlation, 
while the blue circles, representing sources with spectral fits, are more strongly 
anti-correlated.  A partial correlation test shows that the correlation decreases in significance 
from 8.6$\sigma$ ($\tau_{12,3}$ = -0.284) for sources with S/N $>$ 6 to 4.6$\sigma$ 
($\tau_{12,3}$ = -0.131) for sources with S/N $<$ 6.  

A higher percentage of S/N $<$ 6 sources are X-ray weak ($\alpha_{ox}$ $<$ -1.8): 18.7\%, 
compared to only 3.7\% in the SPECTRA (S/N $>$ 6) sample, where absorbed sources have been 
excluded.  Since absorption cannot be accounted for in the low S/N sources, it is a likely 
source of the scatter.  Figure \ref{fig:SNRvsdelgi} confirms that though the SPECTRA sample 
does not remove every X-ray weak source, those with the heaviest optical obscuration 
($\Delta$($g - i$) $\gtrsim$ 0.3) are excluded from the sample.  This illustrates the importance 
of X-ray spectra in excluding absorbed sources.  

For the HIGHSNR sample, the slopes show the same trend as found in the SPECTRA sample (cf. Figures 
4$a$ and 6$a$), though the errors are larger due to the smaller sample size.  Therefore the significant change 
in slope of the $\alpha_{ox}-l_{opt}$ relation is not likely to be due to the effect of undetected absorption 
in the sources with low S/N X-ray spectra.  However, the scatter in the relation defined at low energies does 
decrease for the HIGHSNR sample (cf. Figures 4$c$ and 6$b$), suggesting that undetected absorption affects the 
dispersion of the $\alpha_{ox}-l_{opt}$ relation when sources with low S/N spectra are included.

Results for the CENSORED, SPECTRA and HIGHSNR samples are summarized in Table \ref{tab:results}, which 
gives the slope, intercept and dispersion for the $\alpha_{ox}-l_{opt}$ relation at each X-ray energy 
for each sample.  

\section{Discussion}

The SPECTRA sample will be used for the following discussion.

\subsection{Partial Correlations with Physical Parameters}

To gain increased understanding of the trends found above, we tested for correlations 
with physical parameters.  Controlling for the effects of the optical luminosity, we 
computed the partial correlation coefficient between $\alpha_{ox}$ and black hole mass, 
Eddington ratio, and redshift.  

Black hole masses were derived in \citet{Shen08} using the width and continuum luminosities 
of broad emission lines in the optical spectrum: H$\beta$ for 0 $<$ z $<$ 0.7, MgII for 0.7 
$<$ 1.9, and CIV for 1.9 $<$ z $<$ 4.5.  The broad line region (BLR) is assumed to be 
virialized, so that the continuum luminosities are related to the BLR radius by the R-L 
scaling relation, and the emission line widths give the velocities of the BLR clouds.  
Further details of the line-fitting and scaling relations used for each line can be found 
in \citet{Shen08}.  To calculate the Eddington ratio, we use the \citet{Richards06} 
bolometric correction for 3000 $\mbox{\AA}$, $\kappa_{3000}$ = 5.62$\pm$1.14.  Though 
bolometric corrections are by nature uncertain, Figure 12 of \citet{Richards06} shows that 
3000, as well as 5100 $\mbox{\AA}$ are regions of relatively small dispersion in the 
composite SEDs.  

Tables \ref{tab:aoxMBH}-\ref{tab:aoxz} list the partial correlation coefficients 
($\tau_{12,3}$) and their significance level at each frequency for which $\alpha_{ox}$ 
is defined.  Table \ref{tab:aoxMBH} compares $\alpha_{ox}$ with black hole mass (M$_{BH}$), 
Table \ref{tab:aoxmdot} compares $\alpha_{ox}$ with the Eddington ratio (L$_{bol}$/L$_{Edd}$), 
and Table \ref{tab:aoxz} compares $\alpha_{ox}$ with redshift (z).  We find that 
the primary dependence is of $\alpha_{ox}$ on optical luminosity.  However, there is a 
significant partial correlation with Eddington ratio that increases with X-ray energy, with no 
dependence on optical wavelength (see Fig \ref{fig:aoxEdd}).  The strength of the partial 
correlation suggests that Eddington ratio may play a role in determining $\alpha_{ox}$.

The partial correlation tests show a marginally significant partial correlation between 
$\alpha_{ox}$ and redshift (Table \ref{tab:aoxz}) but the correlation disappears at higher 
energies.  This is likely due to the large scatter in the $\alpha_{ox}-L_{opt}$ relation 
at 1 keV.  The tests show no significant partial correlation between $\alpha_{ox}$ and 
M$_{BH}$ (Table \ref{tab:aoxMBH}).  The black hole masses of the SPECTRA sample range 
from log M$_{BH}$ = 7.4 to 10 M$_{\odot}$.

Note that Kelly et al. (2007; see their Fig. 8)\nocite{Kelly07} found problems with the 
CENS\_TAU program, finding that zero partial correlation between two variables does not 
necessarily result in a null correlation coefficient ($\tau = 0$).  Assessing the real 
significance for a given value of the coefficient is therefore difficult at low values.  
However, given the high value of the partial correlation between $\alpha_{ox}$ and Eddington 
ratio at high X-ray energies ($\tau_{12,3} \sim$ 0.3), a significant correlation appears likely.

\subsection{Consistency Between $\alpha_{ox}-l_{opt}$, $\Gamma-l_x$ and $\Gamma-$L$_{bol}$/L$_{Edd}$ Relations}

In this section, we compare the $\alpha_{ox}-l_{opt}$ relation to the $\Gamma-l_x$ relation 
found in \citet{Young09}.  An apparent contradiction arises, as these relations should lead 
to a secondary relation between $\Gamma$ and $l_{opt}$, when in fact the $\Gamma-l_{opt}$ 
relation is found to be insignificant in the SDSS/XMM-Newton Quasar Survey.  This contradiction 
can be resolved by accounting for the dispersion in the primary relations.  

The relation between $\Gamma$ and $l_x$ \citep{Young09} depends on the X-ray energy.  The 
correlation is marginally significant and positive at the lowest energy ($x$ = 0.7 keV) and 
highly significant and negative at higher energies ($x$ = 4-20 keV).  The change in slope of 
the correlation indicates a pivot in the X-ray spectrum near 1 keV, where the correlation is 
insignificant.  

At 10 keV, the $\Gamma-l_x$ relation is described by
\begin{equation}\Gamma = -(0.335\pm0.032)l_{10keV} + (10.732\pm0.818), 
\sigma = 0.31\label{eqn:Gammalx10},\end{equation}
The $\alpha_{ox}-l_{opt}$ correlation at 2500 $\mbox{\AA}$ and 10 keV is described by
\begin{equation}\alpha_{ox} = -(0.081\pm0.010) l_{2500} + (1.044\pm0.319), 
\sigma = 0.11\label{eqn:aoxlopt10},\end{equation}
which is equivalent to
\begin{equation}l_x = (0.734\pm0.034) l_{2500} + (3.449\pm1.054), 
\sigma = 0.37.\label{eqn:lxlopt10}\end{equation}
Combining equations (\ref{eqn:Gammalx10}) and (\ref{eqn:lxlopt10}) should give the following correlation 
between $\Gamma$ and $l_{opt}$:
\begin{displaymath}\Gamma = -(0.246\pm0.026) l_{2500} + (11.887\pm0.898)\end{displaymath}
Such a correlation is not observed to be significant in the SDSS/XMM-Newton Survey \citep{Young09, 
Risaliti09}, so the relations appear to be inconsistent.  

This apparent contradiction can be explained by accounting for the dispersion associated with the 
$\Gamma - $L$_{bol}$/L$_{Edd}$ relation.  The hard (2-10 keV) X-ray slope has been shown to steepen 
at higher L$_{bol}$/L$_{Edd}$ or at larger FWHM(H$\beta$) \citep{Brandt97,Shemmer06,Kelly08,Risaliti09}.  
\citet{Shemmer08} broke this degeneracy by analyzing a sample of 35 quasars over three orders of 
magnitude in luminosity, finding a stronger $\Gamma-$L$_{bol}$/L$_{Edd}$ than $\Gamma-$FWHM(H$\beta$).  
A linear fit to the SDSS/XMM-Newton data \citep{Risaliti09} gives:
\begin{equation}\Gamma = (0.31\pm0.06) log(L_{bol}/L_{Edd}) + (1.97\pm0.02), 
\sigma = 0.33\label{eqn:GammaLedd}\end{equation}

To illustrate the discussion below, a notional SED is shown in Figure \ref{fig:fakeSED}.  The figure 
shows a simple model of a quasar SED,  where specific luminosities are measured at 2500 $\mbox{\AA}$ 
and 1 keV.  Since the $\Gamma-l_x$ relation is not significant at 1 keV, using the luminosity at this 
energy (instead of at 10 keV, as above) allows the focus to be on the effect of the dispersion associated 
with the $\Gamma-$L$_{bol}$/L$_{Edd}$ relation.  In the figure, model (A) illustrates a source with lower $l_{opt}$ 
(10$^{29}$ ergs s$^{-1}$) and model (B) shows a source with higher $l_{opt}$ (10$^{31}$ ergs s$^{-1}$).  
Inserting the optical luminosity into the appropriate $\alpha_{ox}-l_{opt}$ relations gives an average 
value for $\alpha_{ox}$ (dashed lines) and therefore for $l_{1 keV}$ and $l_{10 keV}$.  

Since 1 keV is near the pivot point of the X-ray spectrum \citep{Young09}, the average X-ray 
slope will not depend on $l_{1 keV}$, and so will just be the mean $\Gamma$ of the sample 
($<\Gamma> \sim$ 1.9).  The dispersion around the mean will arise from two sources: 1) 
dispersion associated with the $\Gamma - $L$_{bol}$/L$_{Edd}$ relation (Eqn \ref{eqn:GammaLedd}, 
shown in Figure \ref{fig:fakeSED}) and 2) dispersion in L$_{bol}$/L$_{Edd}$ for a given 
$l_{opt}$.  While the flattening of the $\alpha_{ox}-l_{opt}$ relation with energy implies 
harder X-ray slope as optical luminosity increases, Fig. \ref{fig:fakeSED} shows that this 
expected hardening is small compared to the dispersion around the mean.  Therefore, any 
secondary relation between $\Gamma-l_{opt}$ would have too large a dispersion for the 
correlation to be significant.  

Moreover, Figure \ref{fig:fakeSED} shows how the $\Gamma-l_x$ relation results naturally as a 
secondary correlation with large scatter.  The figure shows that sources at a given l$_{opt}$ 
with higher hard X-ray luminosities will have harder X-ray slopes, while those with higher soft 
X-ray luminosities will have softer slopes.  The combination of the $\alpha_{ox}-l_{opt}$ 
relation with the $\Gamma$-L$_{bol}$/L$_{Edd}$ relation thus returns the $\Gamma-l_x$ relations 
for X-ray energies $x <$ 1 keV or $x >$ 1 keV.

\subsection{Bolometric Corrections}

As $\alpha_{ox}$ can be used to apply a bolometric correction to the hard X-rays (2-10 keV) 
\citep[e.g., ][]{Hopkins07} by providing a normalization for the X-ray spectrum, we plot in 
Figure \ref{fig:kbol} the dependence of $\kappa_{2-10keV}$ on $\alpha_{ox}$.  Since L$_{bol}$ 
is tightly related to the optical luminosity through the bolometric correction, and 
L$_{2keV}$ is tightly related to L$_{2-10 keV}$ through integration over $\Gamma$, the linear 
correlation is expected.  However, both the correlation and its dispersion, $\sigma$ = 0.12, are of 
interest when determining X-ray bolometric corrections.  For reference, the best-fit line is:
\begin{displaymath} \kappa_{2-10keV} = -(2.282 \pm 0.065)\alpha_{ox} - (1.67 \pm 0.10)\end{displaymath}

The X-ray weak quasars with very large bolometric corrections ($\kappa_{2-10keV}$ $>$ 1000) 
do not have enough S/N for X-ray spectral fits.  One object is optically red ($\Delta$($g-i$) 
= 0.33), so it is possible that these extreme objects are suffering from absorption, affecting 
both their $\alpha_{ox}$ and $\kappa_{2-10keV}$ values.  An additional factor to consider is 
accretion rate.  We find a significant partial correlation between $\kappa_{2-10keV}$ and 
L$_{bol}$/L$_{Edd}$.  The presence of L$_{bol}$ on both axes can result in a false correlation, 
but a partial correlation test with L$_{bol}$ as the third variable shows that the correlation 
is significant at the 7.7$\sigma$ level, with a $\tau_{12,3}$ = 0.232.  The correlation is plotted 
in Figure \ref{fig:kbolEdd} with a best fit line:
\begin{displaymath} log \kappa_{2-10keV} = (0.443 \pm 0.058) log L_{bol}/L_{Edd} + (2.227 \pm 0.045)\end{displaymath}
This correlation is consistent with previous results \citep[e.g.,][]{VF09}, and suggests 
that the fraction of disk photons up-scattered to the X-rays decreases as accretion rate increases.

\section{Conclusions}

We find that the slope of the $\alpha_{ox}-l_{opt}$ relation depends on the X-ray energy at which 
$\alpha_{ox}$ is defined, though there is no significant dependence on optical wavelength.  While 
a change in slope is predicted when the baseline over which $\alpha_{ox}$ is defined increases or 
decreases, the slopes at the lowest X-ray energy are significantly steeper than predicted, while the 
slopes at high X-ray energies are marginally flatter than predicted.  This suggests that the 
efficiency of low energy X-ray photon production depends more strongly on the optical/UV photon 
supply, whereas the efficiency of high energy X-ray photon production remains relatively constant 
with respect to the seed photon supply.  Partial correlation tests show that Eddington ratio is highly 
correlated with $\alpha_{ox}$ when controlling for $l_{opt}$, and this partial correlation increases 
in significance with X-ray energy, suggesting that accretion rate likely plays a role in the results.  
Nevertheless, $l_{opt}$ remains the primary driver of $\alpha_{ox}$.  

To check the results, we redo the calculations for a sample including low S/N sources (CENSORED) and for a 
sample including only high S/N sources (HIGHSNR).  The CENSORED sample shows the same trends as in the 
SPECTRA sample, though the results are less significant.  This is likely due 
to the diluting effect of absorption in the low S/N sources, which emphasizes the importance of X-ray 
spectra in eliminating sources with absorption.  The HIGHSNR sample reproduces the trend in the slope of 
the $\alpha_{ox}-l_{opt}$ relation found for the SPECTRA sample, indicating that the results are not 
affected by undetected absorption.  The dispersion of the relation does decrease significantly for the 
HIGHSNR sample, suggesting that undetected absorption does introduce some error into the determination 
of X-ray fluxes for sources with moderate S/N.  

By combining the $\alpha_{ox}-l_{opt}$ relation with the $\Gamma$-L$_{bol}$/L$_{Edd}$ relation, 
we can naturally explain two results: 1) the existence of the $\Gamma-l_x$ relation as reported 
in \citet{Young09} and 2) the lack of a $\Gamma-l_{opt}$ relation.  Given the measured dispersions 
and correlation coefficients, it is likely that the $\alpha_{ox}-l_{opt}$ and $\Gamma$-
L$_{bol}$/L$_{Edd}$ relations are the primary relations, while the $\Gamma-l_x$ relation is 
secondary.  Understanding the relations in this order will give a more complete framework for 
understanding the relation between optical and X-ray emission in quasars.  In order to fully 
understand the physical implications of the results shown here, a study of the selection, 
orientation and variability effects is needed.  Such a study is now in progress.  

Finally, for the purposes of aiding studies of the hard X-ray bolometric correction, we 
include a description of the $\kappa_{2-10keV}$ - $\alpha_{ox}$ correlation and dispersion.  We 
find a significant partial correlation between $\kappa_{2-10keV}$ and L$_{bol}$/L$_{Edd}$, when 
taking bolometric luminosity into account as the third variable.  This is consistent with a 
decreasing fraction of up-scattered X-ray photons as accretion rate increases.

\acknowledgements

M.Y. thanks Ranjan Vasudevan for helpful conversations regarding this work.  
We thank the referee for useful comments that improved the quality of this paper.
This paper is based on observations obtained with XMM-Newton, an ESA science mission with 
instruments and contributions directly funded by ESA Member States and NASA, and the Sloan 
Digital Sky Survey (SDSS).  Funding for the SDSS and SDSS-II has been provided by the 
Alfred P.  Sloan Foundation, the Participating Institutions, the National Science Foundation, 
the U.S.  Department of Energy, the National Aeronautics and Space Administration, the 
Japanese Monbukagakusho, the Max Planck Society, and the Higher Education Funding Council 
for England.  This research also made use of the NASA/ IPAC Infrared Science Archive, 
which is operated by the Jet Propulsion Laboratory, California Institute of Technology, 
under contract with the National Aeronautics and Space Administration.  This work has 
been partially funded by NASA Grants NASA NNX07AI22G and NASA GO6-7102X. 



\begin{deluxetable}{lllllll}
\tabletypesize{\footnotesize}
\tablecolumns{7} 
\tablewidth{0pt} 
\tablecaption{Sample Definitions\label{tab:samples}}
\tablehead{\colhead{Sample name}				&
	   \colhead{N$_{tot}$}		              		&
	   \colhead{S/N} 				 	&
	   \colhead{median N$_H$}			 	&
	   \colhead{z}					 	&
	   \colhead{$i$ mag}				 	&
	   \colhead{\% targets}		 		 	}
\startdata
CENSORED	&  514  &   all  	&  0.39  &  0.1-5.0  &  15.2-20.7  & $\sim$2\% \\
SPECTRA		&  327  &   S/N $>$ 6   &  0.39  &  0.1-4.4  &  15.2-20.4  & $\sim$3\% \\
HIGHSNR		&  99   &   S/N $>$ 20  &  0.06  &  0.1-3.0  &  15.6-19.9  & $\sim$9\% \\
\enddata
\tablecomments{All three samples exclude RL and BAL quasars, sources 
with significant intrinsic absorption, and sources with bad fits to 
the X-ray spectrum.}
\end{deluxetable}

\begin{deluxetable}{lllllllllllllllllll}
\tabletypesize{\footnotesize}
\tablecolumns{13} 
\tablewidth{0pt} 
\tablecaption{Results for the $\alpha_{ox}-l_{opt} Relation$\label{tab:results}}
\tablehead{\colhead{X-ray Energy}				&
	   \colhead{}			              		&
	   \colhead{CENSORED} 				 	&
	   \colhead{}						&
	   \colhead{}			              		&
	   \colhead{SPECTRA} 				 	&
	   \colhead{}						&
	   \colhead{}			              		&
	   \colhead{HIGHSNR} 				 	&
	   \colhead{}						\\
	   \colhead{}						&
	   \colhead{Slope}					&
	   \colhead{Intercept}					&
	   \colhead{$\sigma$}					&
	   \colhead{Slope}					&
	   \colhead{Intercept}					&
	   \colhead{$\sigma$} 				&
	   \colhead{Slope}					&
	   \colhead{Intercept}					&
	   \colhead{$\sigma$}					}
\startdata
1 keV   & -0.17$\pm$0.02  & 3.47$\pm$0.47 & 0.19 & -0.19$\pm$0.02  & 4.13$\pm$0.46 & 0.15 & -0.13$\pm$0.02 & 2.45$\pm$0.54 & 0.11 \\
1.5 keV & -0.13$\pm$0.01  & 2.43$\pm$0.42 & 0.17 & -0.14$\pm$0.01  & 2.70$\pm$0.38 & 0.13 & -0.10$\pm$0.02 & 1.55$\pm$0.50 & 0.10 \\
2 keV   & -0.12$\pm$0.01  & 2.13$\pm$0.39 & 0.16 & -0.12$\pm$0.01  & 2.27$\pm$0.35 & 0.11 & -0.08$\pm$.02 & 1.09$\pm$0.49 & 0.10 \\
4 keV   & -0.099$\pm$0.01 & 1.49$\pm$0.35 & 0.14 & -0.098$\pm$0.01 & 1.54$\pm$0.30 & 0.10 & -0.06$\pm$0.02 & 0.30$\pm$0.53 & 0.11 \\
7 keV   & -0.090$\pm$0.01 & 1.26$\pm$0.34 & 0.14 & -0.086$\pm$0.01 & 1.21$\pm$0.31 & 0.10 & -0.04$\pm$0.02 &-0.15$\pm$0.58 & 0.12 \\
10 keV  & -0.080$\pm$0.01 & 0.97$\pm$0.34 & 0.13 & -0.081$\pm$0.01 & 1.04$\pm$0.32 & 0.11 & -0.03$\pm$0.02 &-0.40$\pm$0.61 & 0.12 \\
\enddata
\tablecomments{Slopes, intercepts, and dispersions for each of the three samples at each X-ray energy, 
using 2500 $\mbox{\AA}$ as the optical wavelength.  Errors are at the 1$\sigma$ level.}
\end{deluxetable}

\begin{deluxetable}{llll}
\tabletypesize{\footnotesize}
\tablecolumns{4} 
\tablewidth{0pt} 
\tablecaption{$\alpha_{ox}$ - M$_{BH}$ Partial Correlation Coefficients\label{tab:aoxMBH}}
\tablehead{\colhead{X-ray Energy (keV)}				&
	   \colhead{1500 $\mbox{\AA}$}              		&
	   \colhead{2500 $\mbox{\AA}$} 		 	&
	   \colhead{5000 $\mbox{\AA}$} 		 	}
\startdata
1.0	& -0.070 (1.84) & -0.064 (1.68)  & -0.065 (1.71) \\
1.5	& -0.036 (0.97) & -0.030 (0.81)  & -0.032 (0.86) \\
2.0	& -0.013 (0.35) & -0.0071 (0.20) & -0.0093 (0.26) \\
4.0	& 0.048 (1.37)  & 0.053 (1.51)   & 0.047 (1.38) \\
7.0	& 0.093 (2.66)  & 0.096 (2.82)   & 0.086 (2.53) \\
10.0	& 0.11 (3.14)   & 0.10 (2.94)    & 0.10 (2.94) \\
\enddata
\tablecomments{The partial correlation coefficient listed is Kendall's $\tau_{12,3}$, with its significance 
($\sigma$) in parantheses.  Correlations with significance above 3.3$\sigma$ (P $>$ 99.9\%) are considered 
to be significant.  These are the results for the SPECTRA sample. }
\end{deluxetable}

\begin{deluxetable}{llll}
\tabletypesize{\footnotesize}
\tablecolumns{4} 
\tablewidth{0pt} 
\tablecaption{$\alpha_{ox}$ - L$_{bol}$/L$_{Edd}$ Partial Correlation Coefficients\label{tab:aoxmdot}}
\tablehead{\colhead{X-ray Energy (keV)}				&
	   \colhead{1500 $\mbox{\AA}$}              		&
	   \colhead{2500 $\mbox{\AA}$} 		 	&
	   \colhead{5000 $\mbox{\AA}$} 		 	}
\startdata
1.0	& -0.13 (3.25) & -0.12 (3.00) & -0.12 (3.00) \\
1.5	& -0.17 (4.25) & -0.16 (4.00) & -0.16 (4.00) \\
2.0	& -0.19 (4.87) & -0.19 (4.87) & -0.19 (4.87) \\
4.0	& -0.27 (7.11) & -0.27 (7.11) & -0.27 (7.11) \\
7.0	& -0.32 (8.21) & -0.32 (8.21) & -0.32 (8.21) \\
10.0	& -0.33 (8.25) & -0.33 (8.25) & -0.33 (8.25) \\
\enddata
\tablecomments{The partial correlation coefficient listed is Kendall's $\tau_{12,3}$, with its significance 
($\sigma$) in parantheses.  Correlations with significance above 3.3$\sigma$ (P $>$ 99.9\%) are considered 
to be significant.  These are the results for the SPECTRA sample.}
\end{deluxetable}

\begin{deluxetable}{llll}
\tabletypesize{\footnotesize}
\tablecolumns{4} 
\tablewidth{0pt} 
\tablecaption{$\alpha_{ox}$ - z Partial Correlation Coefficients\label{tab:aoxz}}
\tablehead{\colhead{X-ray Energy (keV)}				&
	   \colhead{1500 $\mbox{\AA}$}              		&
	   \colhead{2500 $\mbox{\AA}$} 		 	&
	   \colhead{5000 $\mbox{\AA}$} 		 	}
\startdata
1.0	& -0.16 (4.32)  & -0.15 (4.05)  & -0.14 (0.036) \\
1.5	& -0.11 (3.14)  & -0.097 (2.77) & -0.087 (2.56) \\
2.0	& -0.090 (2.65) & -0.074 (2.18) & -0.064 (1.88) \\
4.0	& -0.040 (1.21) & -0.027 (0.84) & -0.021 (0.66) \\
7.0	& -0.022 (0.67) & -0.011 (0.33) & -0.010 (0.31) \\
10.0	& -0.015 (0.45) & -0.013 (0.41) & 0.0086 (0.26) \\
\enddata
\tablecomments{The partial correlation coefficient listed is Kendall's $\tau_{12,3}$, with its significance 
($\sigma$) in parantheses.  Correlations with significance above 3.3$\sigma$ (P $>$ 99.9\%) are considered 
to be significant.  These are the results for the SPECTRA sample.}
\end{deluxetable}

\begin{figure}
\centering
\includegraphics[width=3.2in]{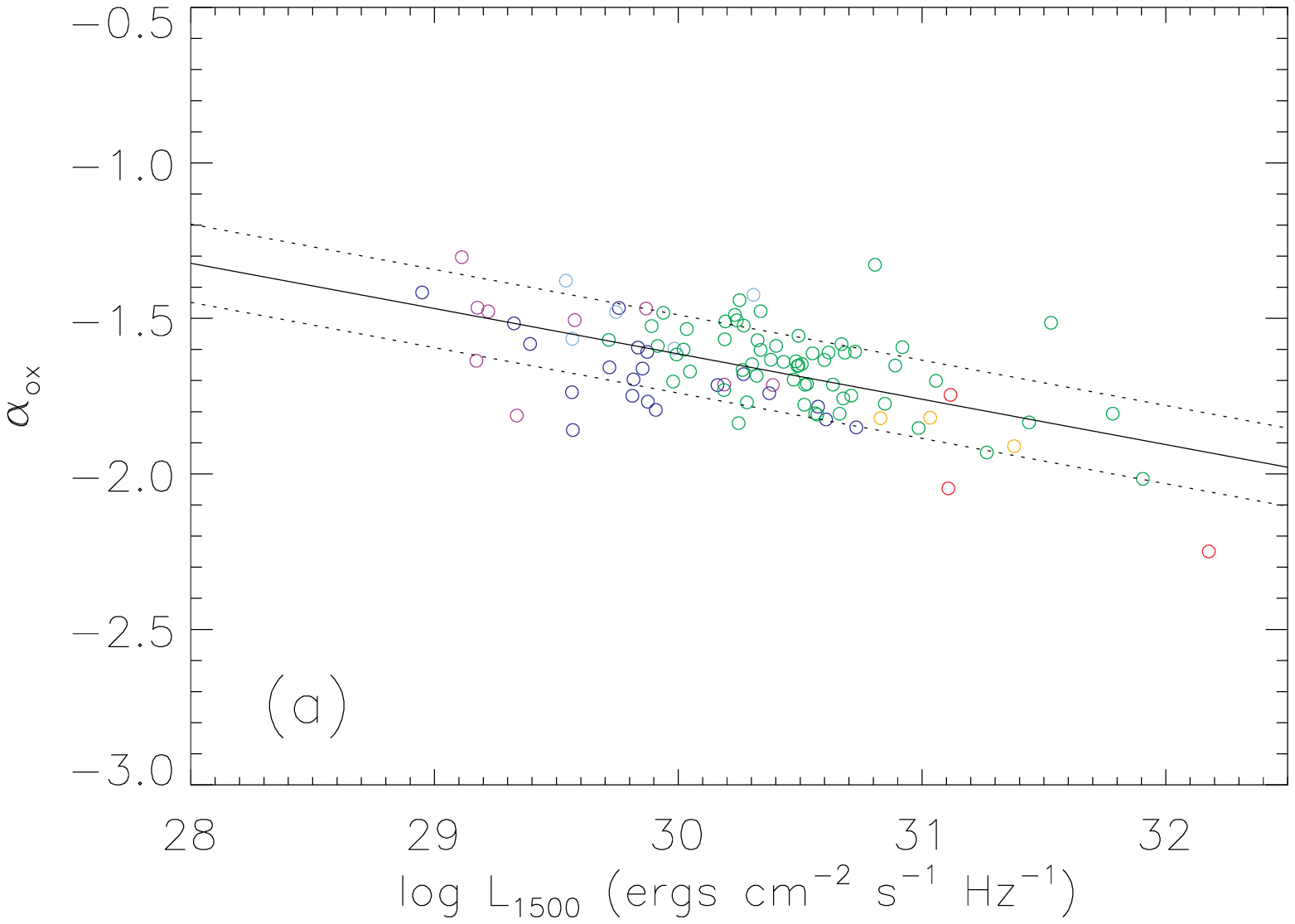}
\includegraphics[width=3.2in]{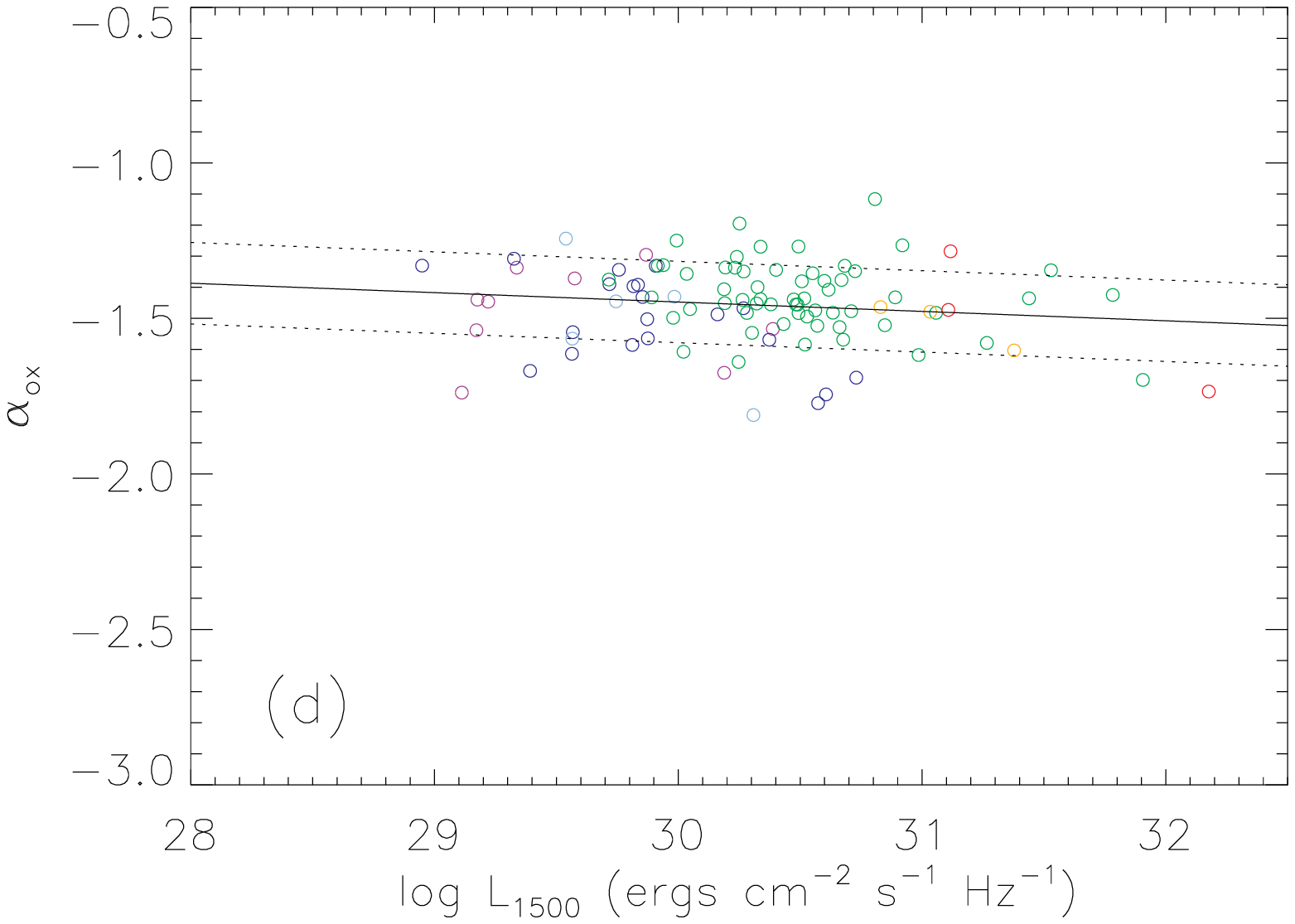}
\includegraphics[width=3.2in]{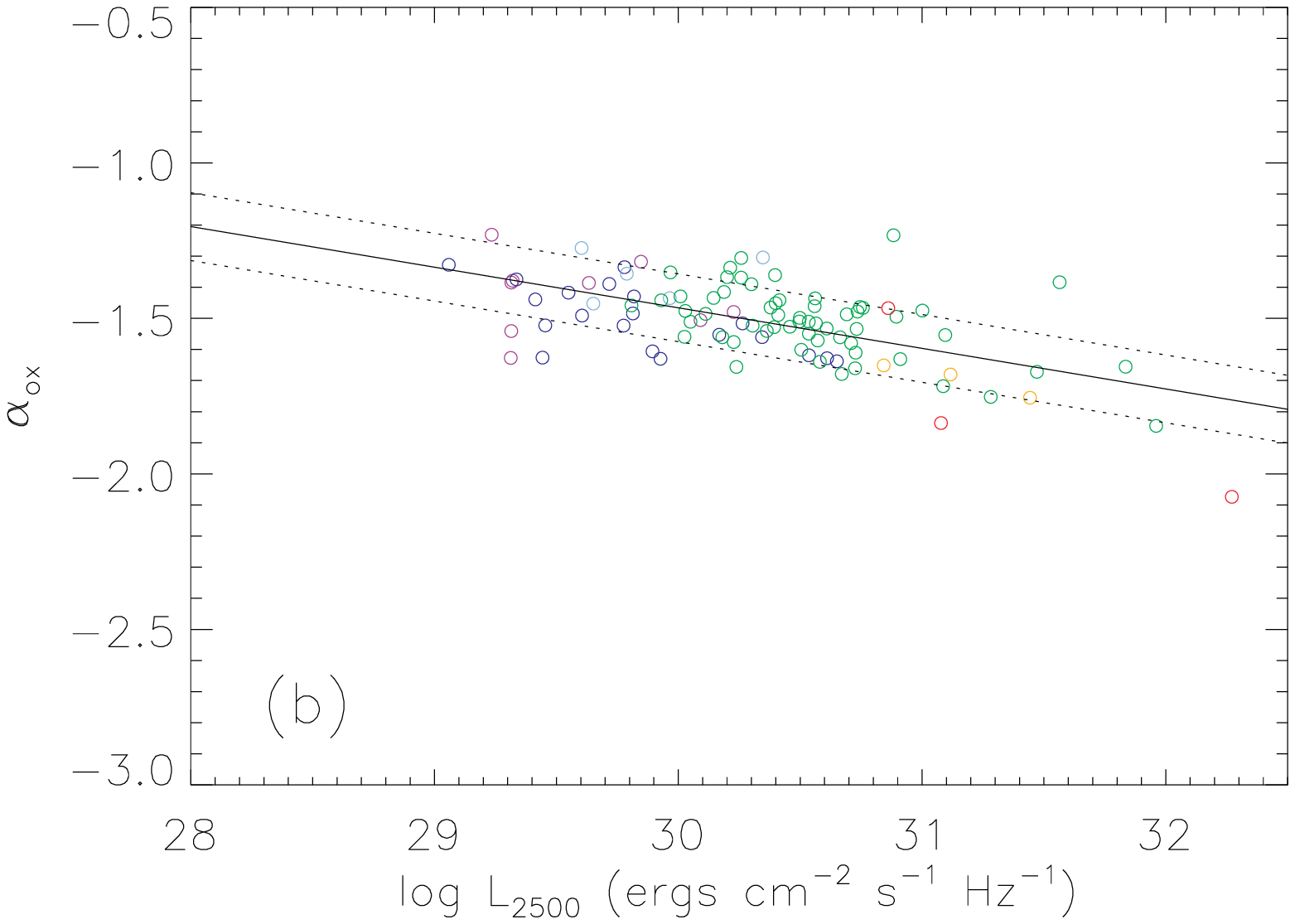}
\includegraphics[width=3.2in]{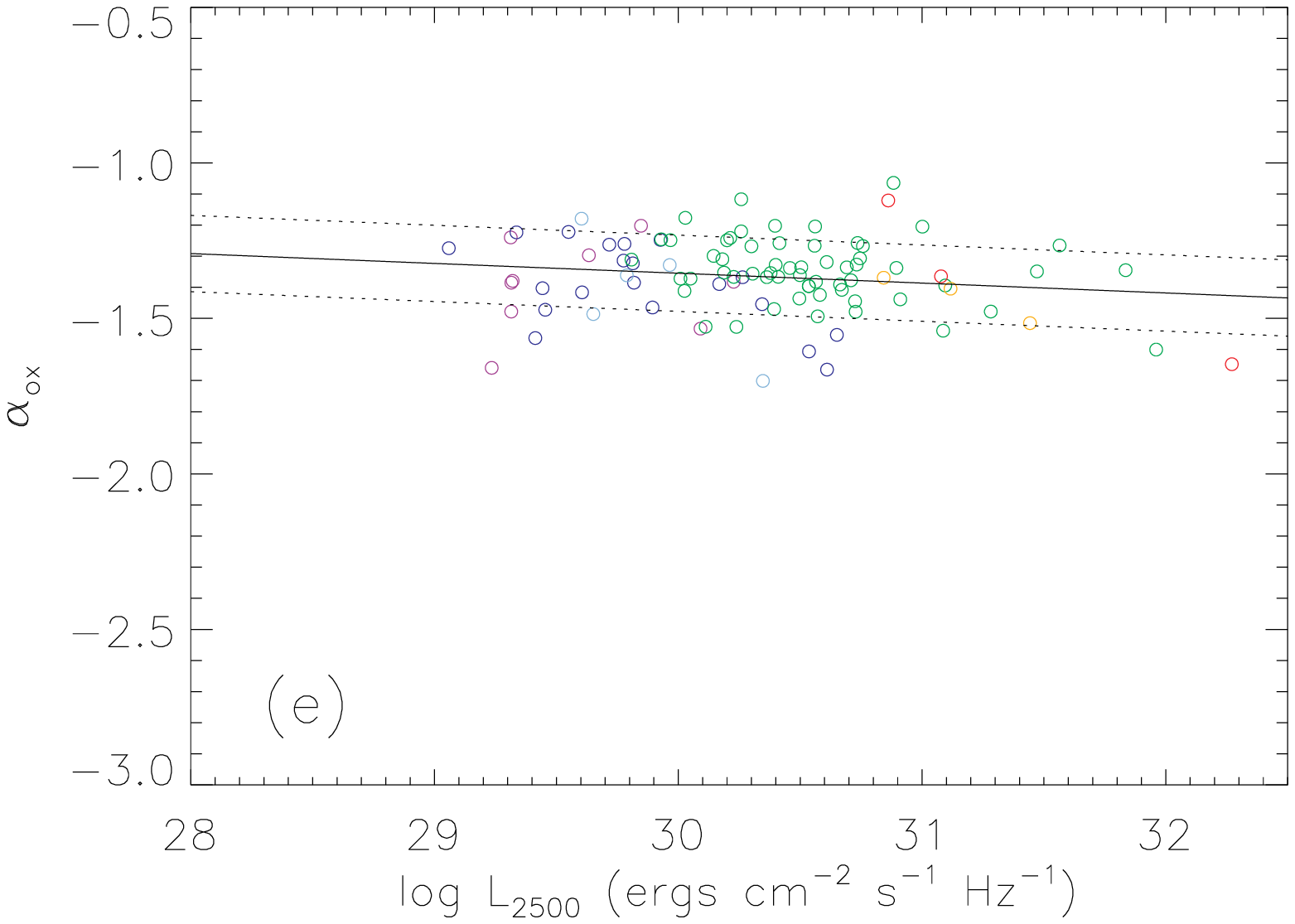}
\includegraphics[width=3.2in]{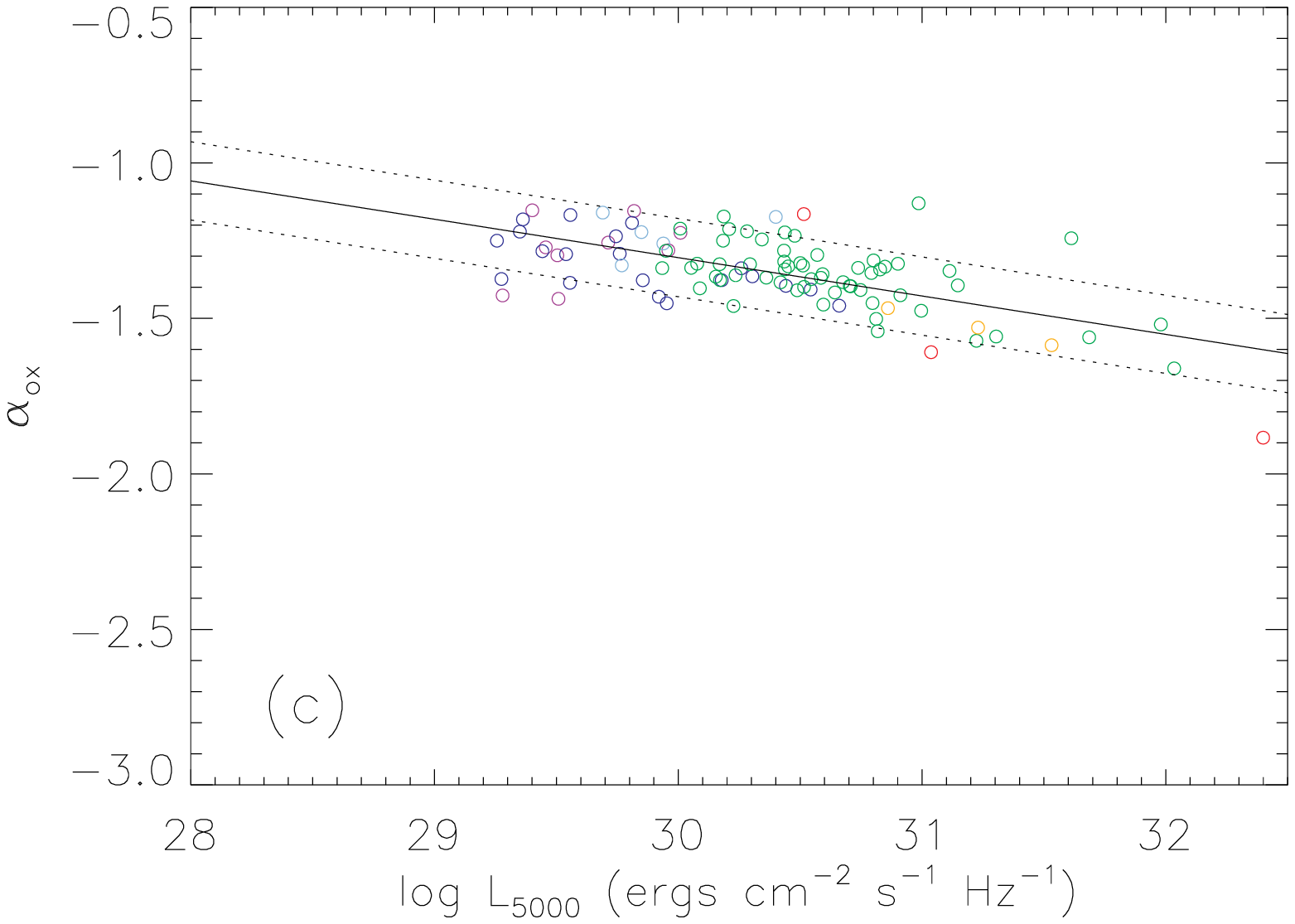}
\includegraphics[width=3.2in]{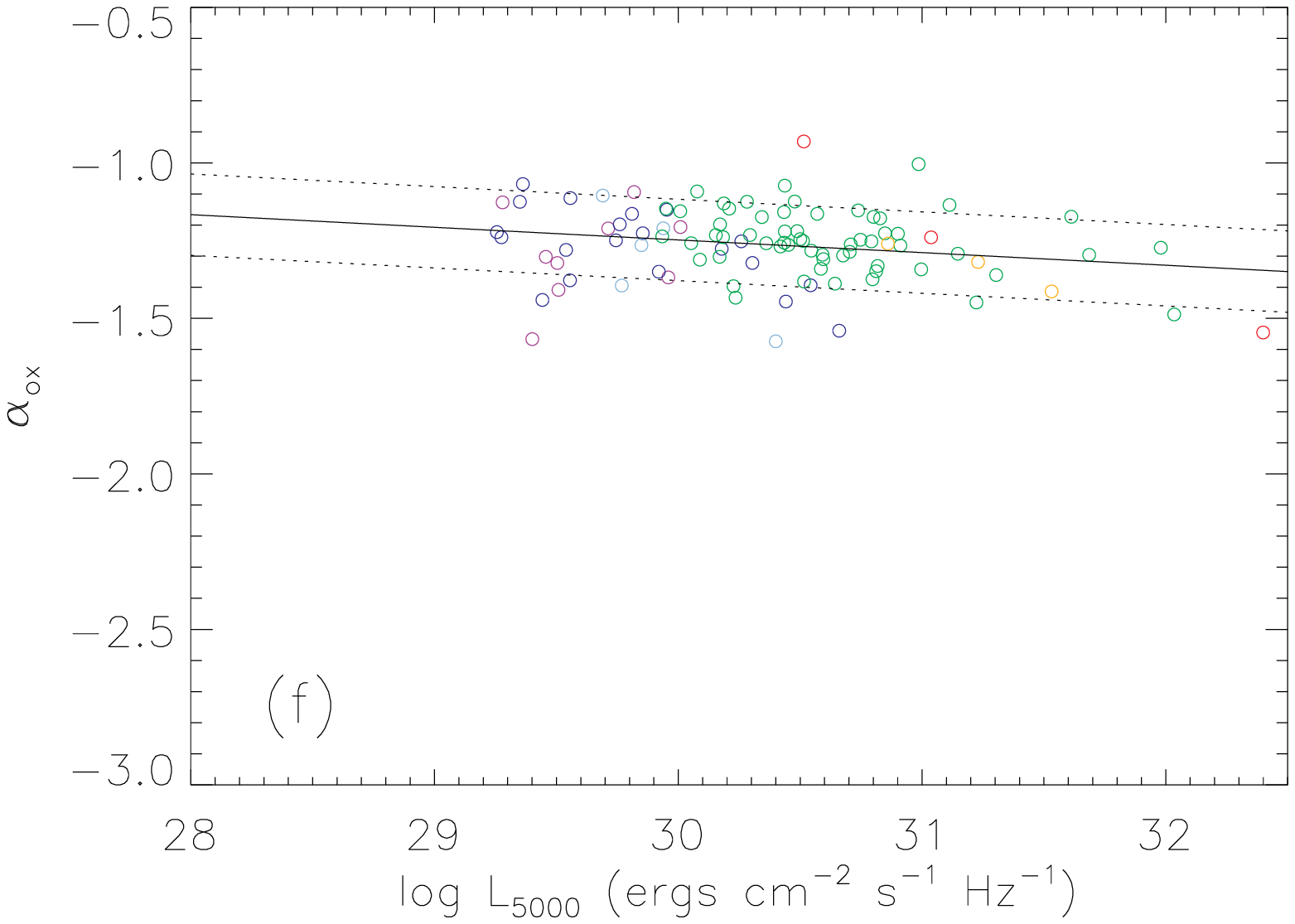}
\caption{(a)-(c) The $\alpha_{ox}-l_{opt}$ relation for the SPECTRA sample, where the X-ray energy 
is 1 keV and the optical luminosities are taken at 1500, 2500 and 5000 $\mbox{\AA}$.  
(d)-(f) The $\alpha_{ox}-l_{opt}$ relation, where the X-ray energy is 10 keV and the optical 
luminosities are taken at 1500, 2500 and 5000 $\mbox{\AA}$.  Only sources with S/N $>$ 6 are plotted.  
The colors signify redshift bins, which were used in determining the optical luminosities.  
Purple (z $<$ 0.25), blue (0.25 $<$ z $<$ 0.43), light blue (0.43 $<$ z $<$ 0.55), 
green (0.55 $<$ z $<$ 1.8), orange (1.8 $<$ z $<$ 2.3), red (2.3 $<$ z $<$ 3.75) and black (z $>$ 3.75).  
BAL and RL quasars are excluded, as are quasars with bad X-ray fits and quasars with X-ray fits 
that require significant intrinsic absorption.  }
\label{fig:aoxvsLopt_new}
\end{figure}

\begin{figure}
\centering
\includegraphics[width=3.2in]{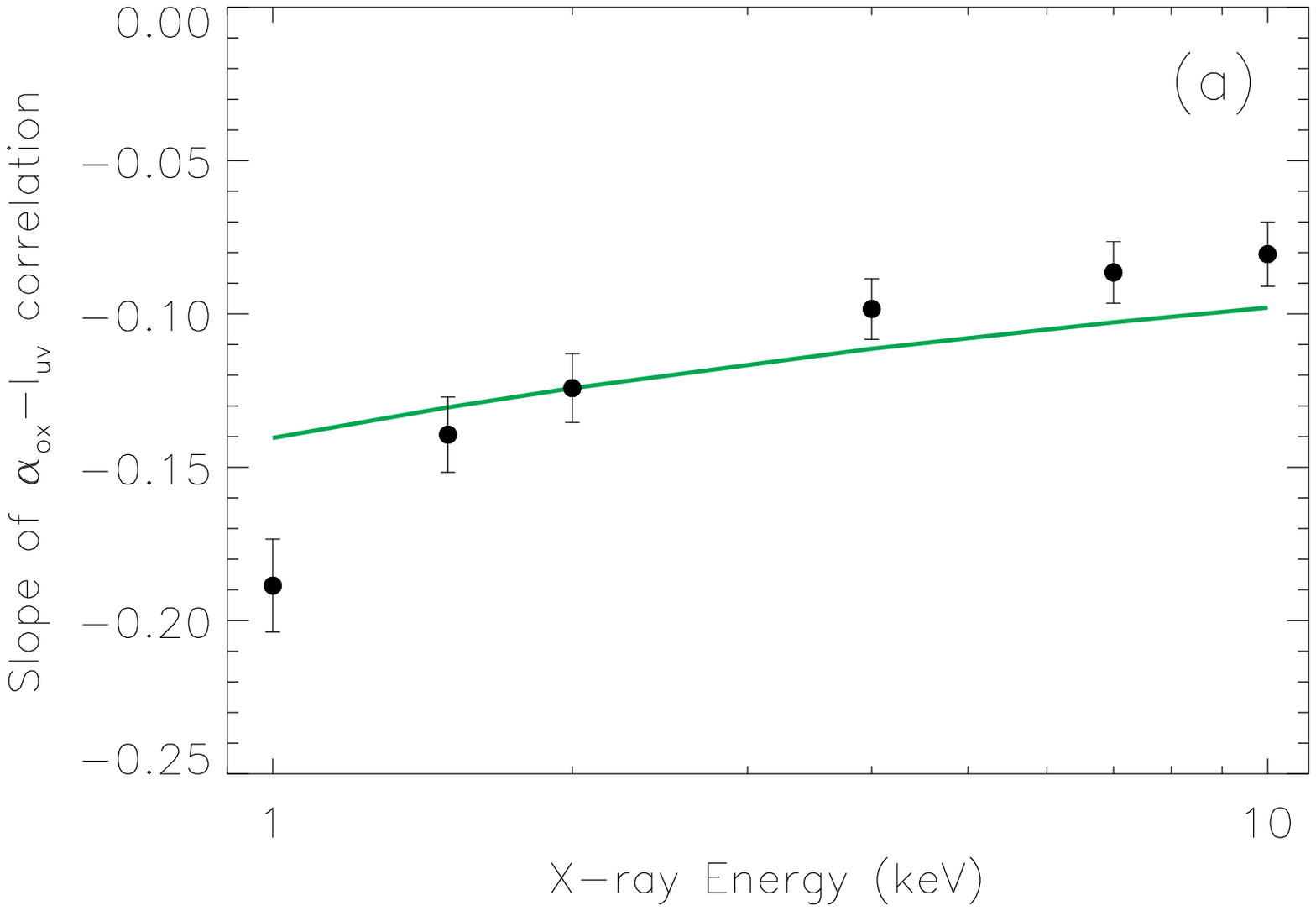}
\includegraphics[width=3.2in]{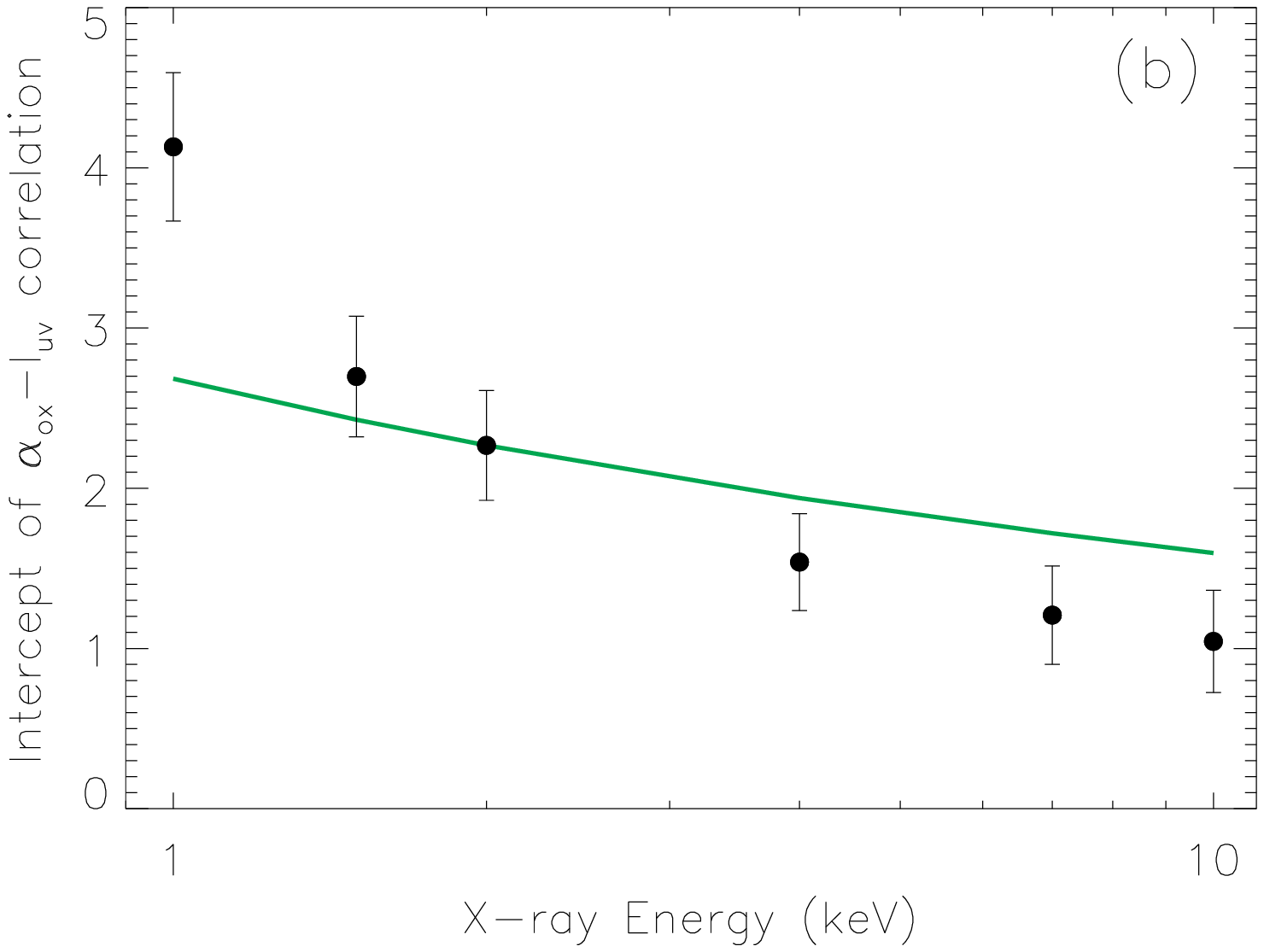}
\includegraphics[width=3.2in]{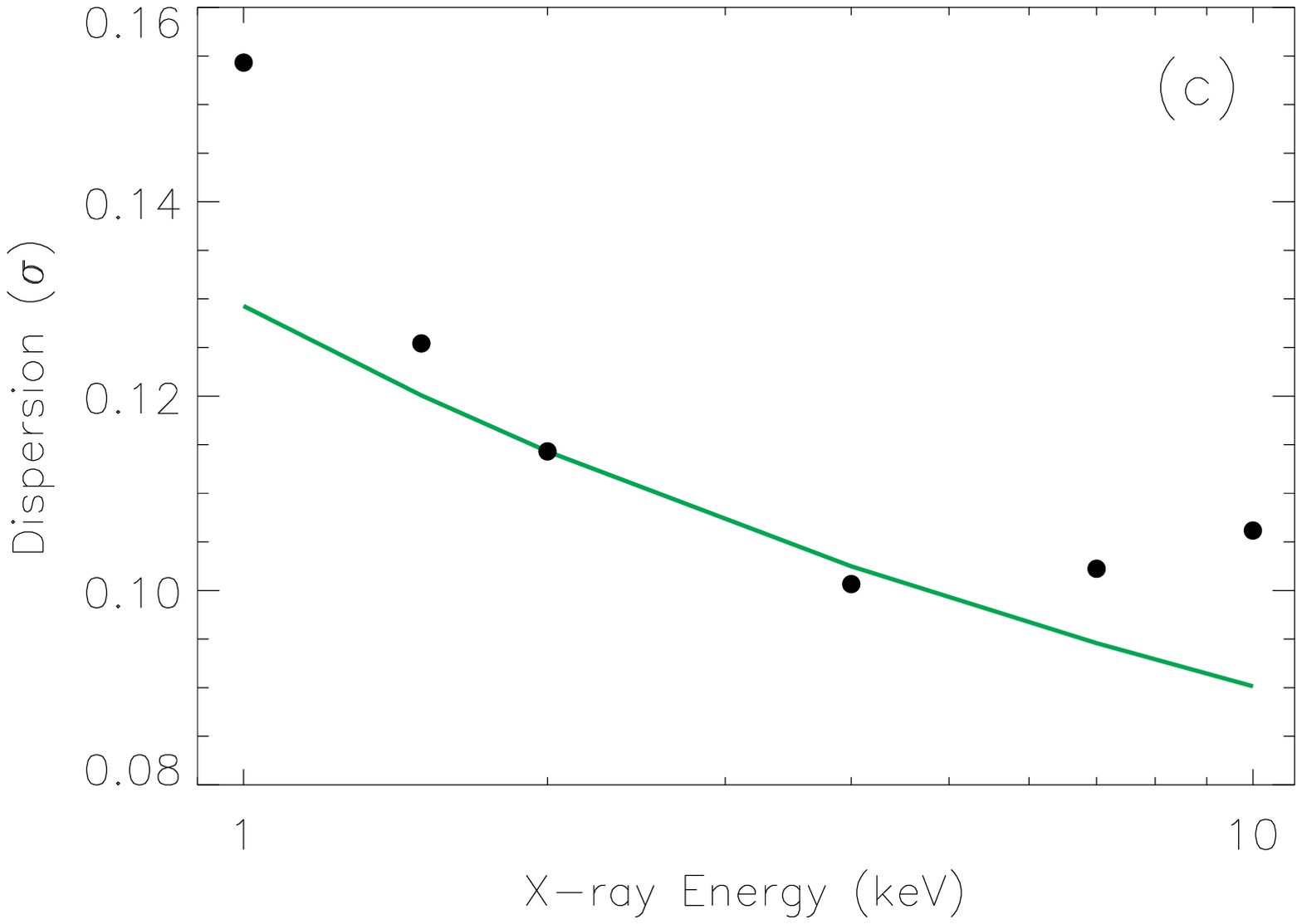}
\caption{The slope (a), intercept (b) and dispersion (c) of the $\alpha_{ox}-l_{opt}$ relation 
with respect to the X-ray energy for the SPECTRA sample.  The optical wavelength is held constant at 2500 
$\mbox{\AA}$.  The filled 
circles show the measured values, and the solid line shows the calculated values.  Values are calculated by 
normalizing at the slope/intercept/dispersion for the relation at 2500 $\mbox{\AA}$ and 2 keV, and then 
correcting for the baseline over which $\alpha_{ox}$ is defined.  Measured values that lie away from the 
calculated baseline effect are affected by measurement errors and/or an intrinsic change in the 
$\alpha_{ox}-l_{opt}$ relation.}
\label{fig:SPECTRA}
\end{figure}

\begin{figure}
\centering
\includegraphics[width=3.2in]{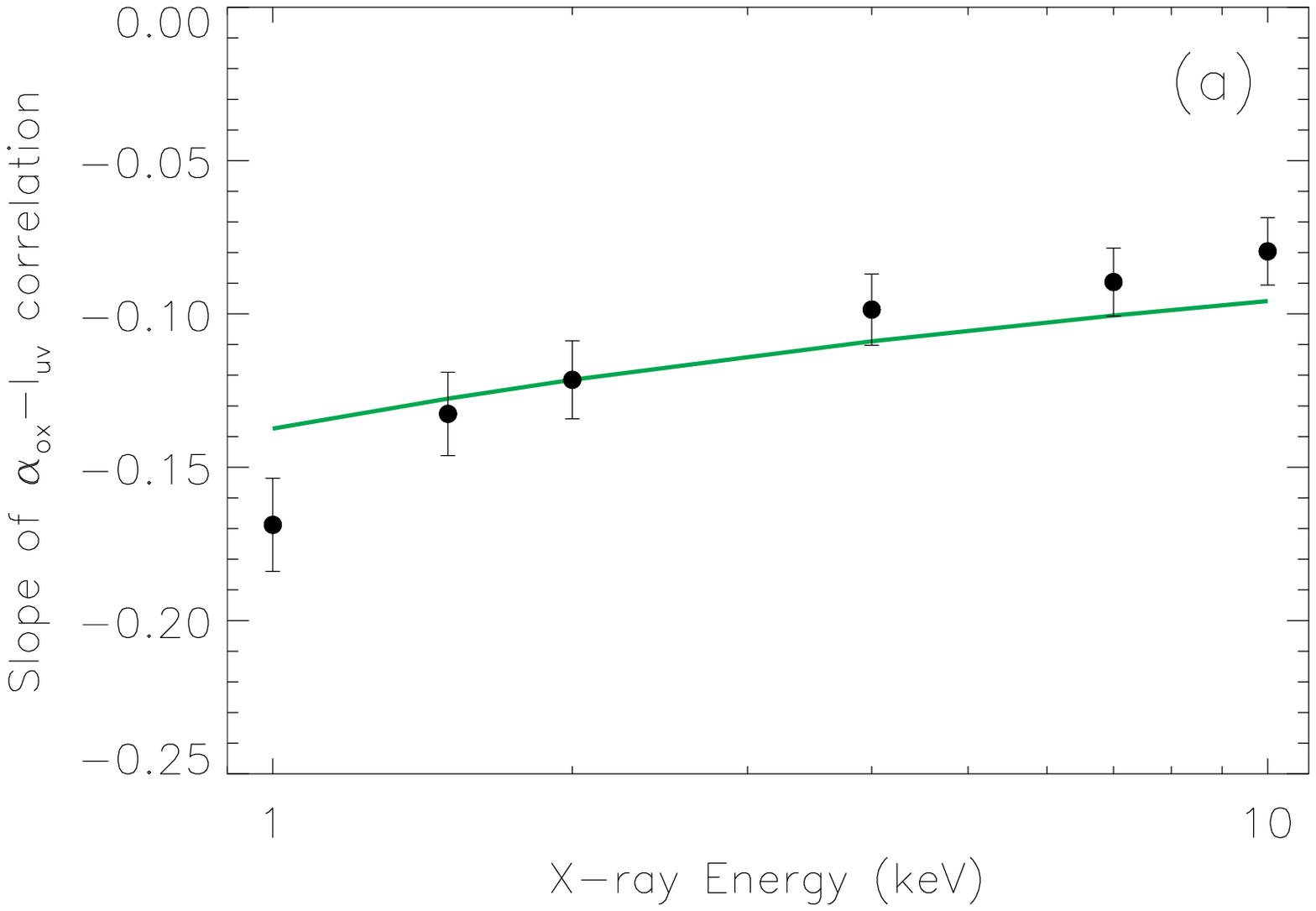}
\includegraphics[width=3.2in]{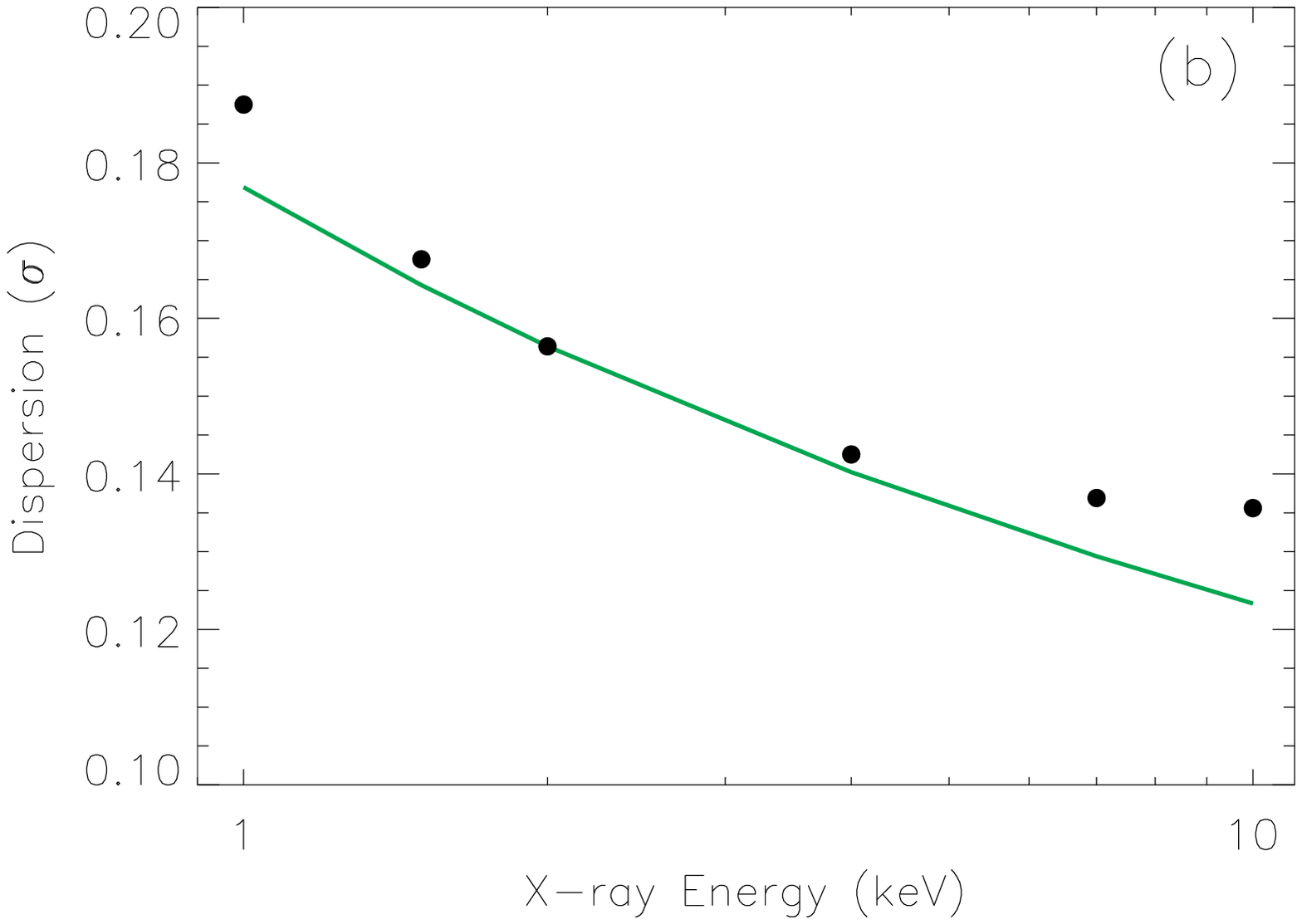}
\caption{The slope (a) and dispersion (b) of the $\alpha_{ox}-l_{opt}$ relation for the CENSORED 
sample, as in Figure 3.}
\label{fig:CENSORED}
\end{figure}

\begin{figure}
\centering
\includegraphics[]{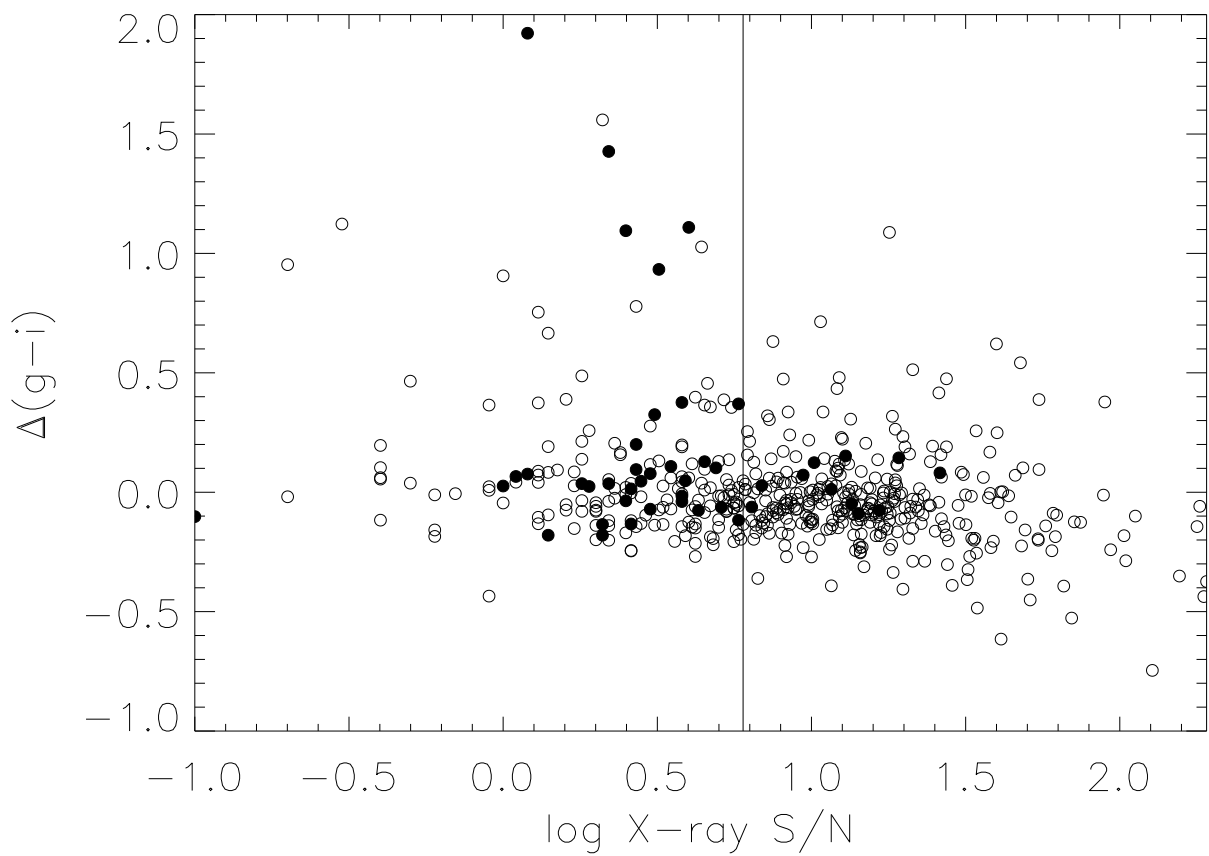}
\caption{Optical color $\Delta$($g - i$) vs. X-ray S/N for the CENSORED sample.  The SPECTRA 
sample contains sources with S/N $>$ 6, marked with a solid vertical line.  X-ray weak sources 
($\alpha_{ox}$ $<$ -1.8) are marked as solid circles.}
\label{fig:SNRvsdelgi}
\end{figure}

\begin{figure}
\centering
\includegraphics[width=3.2in]{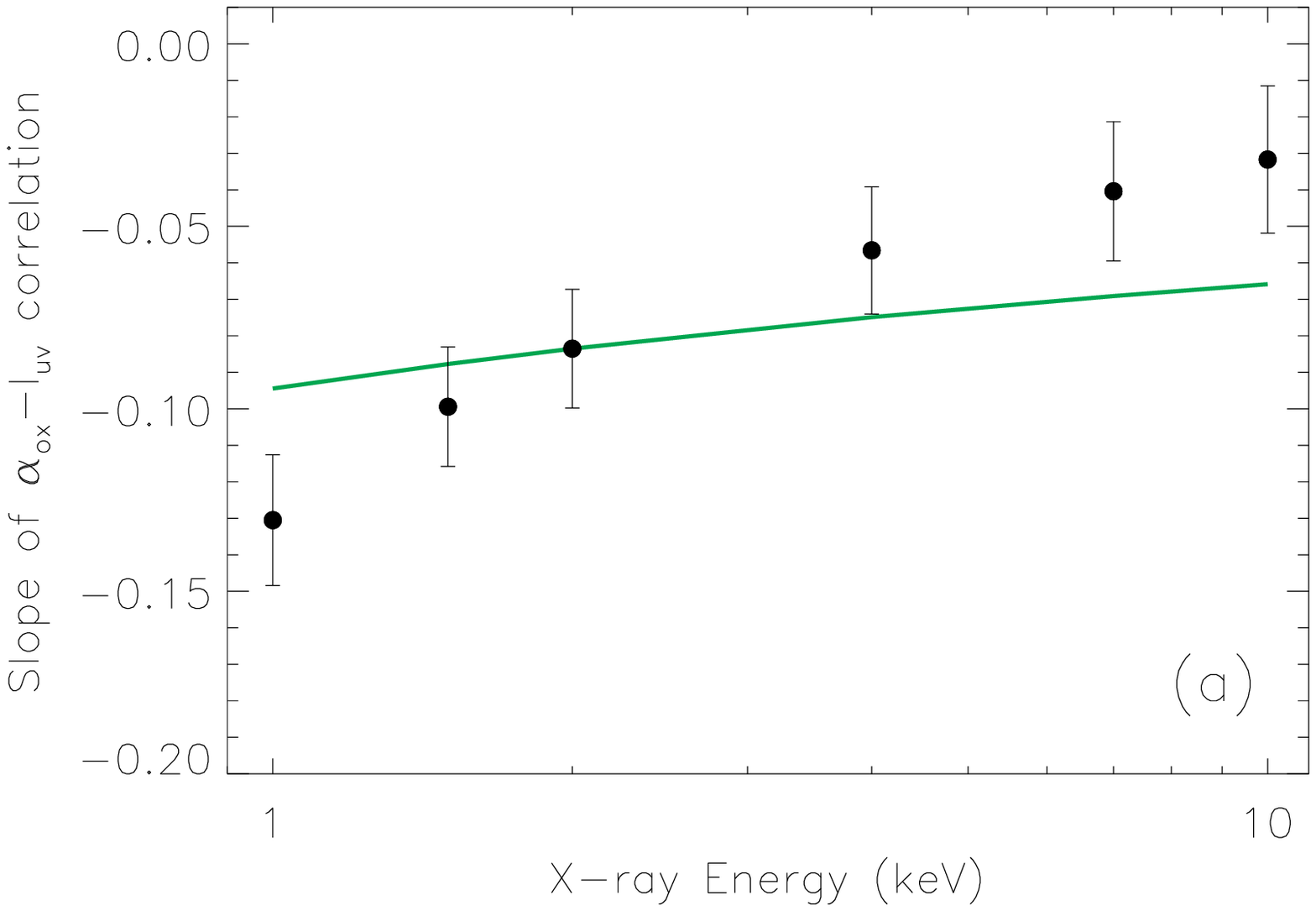}
\includegraphics[width=3.2in]{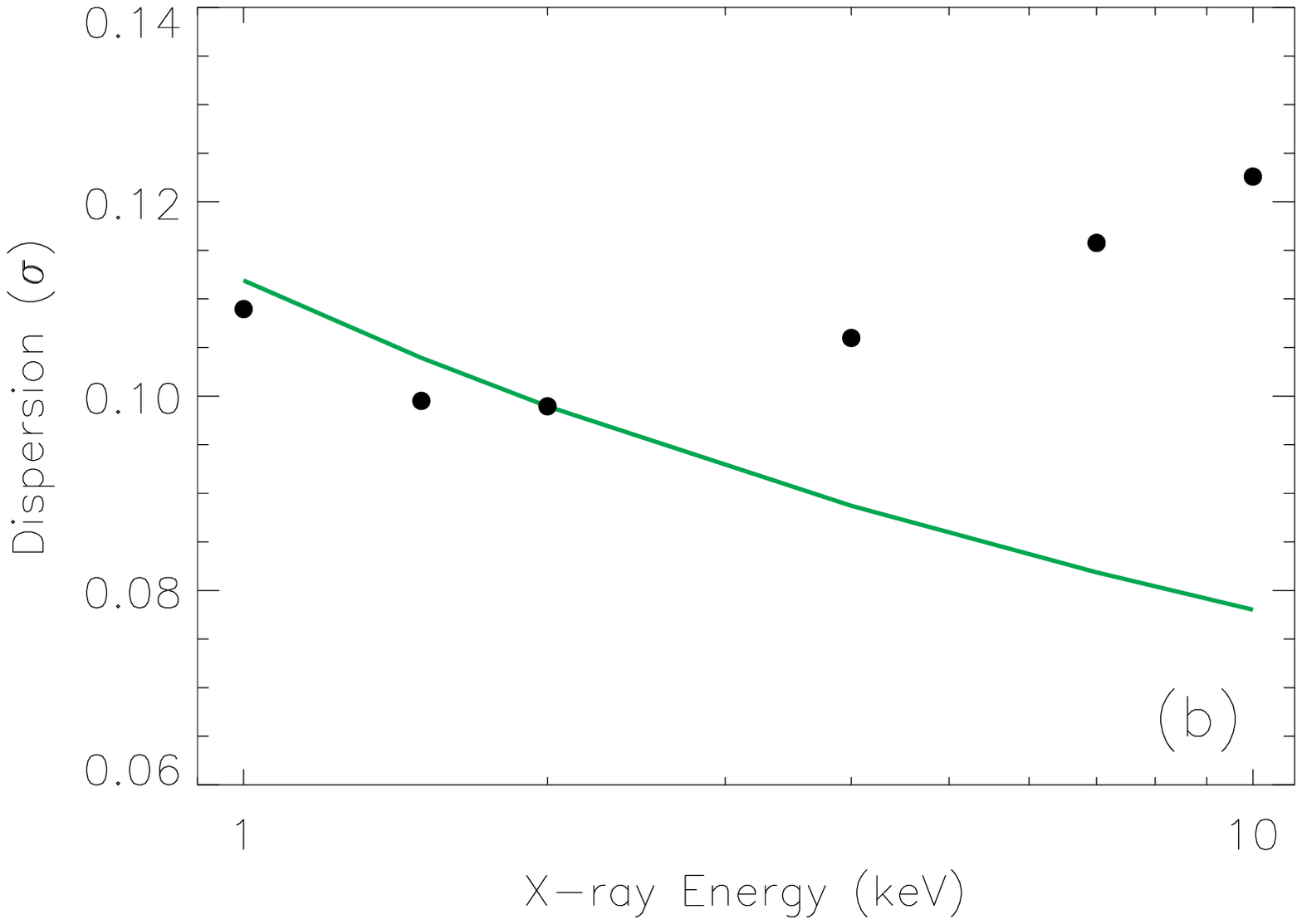}
\caption{The slope (a) and dispersion (b) of the $\alpha_{ox}-l_{opt}$ relation for the HIGHSNR 
sample, as in Figure 3.}
\label{fig:HIGHSNR}
\end{figure}


\begin{figure}
\centering
\includegraphics[]{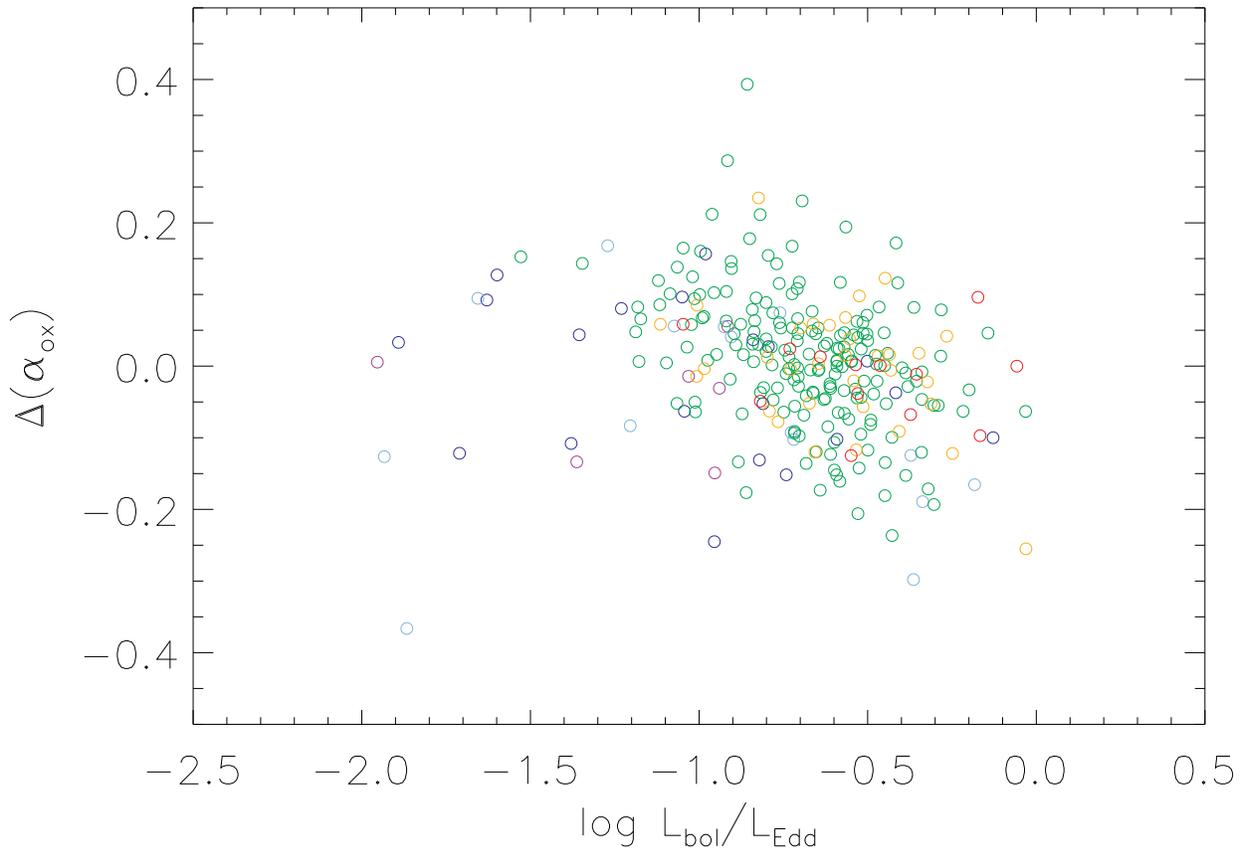}
\caption{$\Delta\alpha_{ox}$ vs. L$_{bol}$/L$_{Edd}$.  Defined at 4 keV and 2500 $\mbox{\AA}$, 
$\Delta\alpha_{ox}$ is the difference between the calculated $\alpha_{ox}$ and that predicted 
by the relevant $\alpha_{ox}-l_{opt}$ relation.  Colors show redshift ranges, as in Fig. 3.}
\label{fig:aoxEdd}
\end{figure}

\begin{figure}
\centering
\includegraphics[]{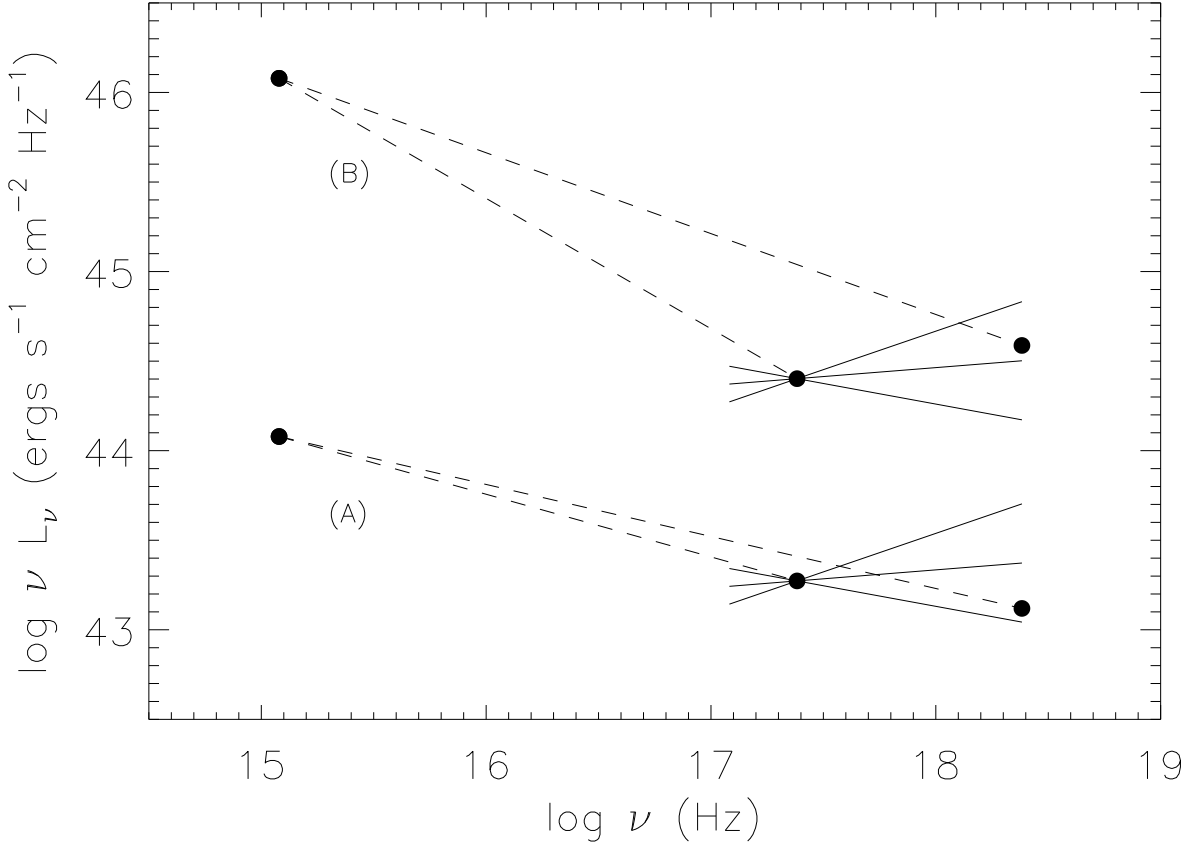}
\caption{Two notional quasar SEDs are shown, where specific luminosities are measured at 
2500 $\mbox{\AA}$, 1 keV and 10 keV.  Model (A) illustrates a source with lower $l_{opt}$ 
(10$^{29}$ ergs s$^{-1}$) and (B) a source with higher $l_{opt}$ (10$^{31}$ ergs s$^{-1}$).  
The X-ray luminosities ($l_{1 keV}$, $l_{10 keV}$) are determined from the appropriate 
$\alpha_{ox}-l_{opt}$ relations, and the resulting imaginary power-laws defined by 
$\alpha_{ox}$ are shown as dashed lines.  The average X-ray photon index ($<\Gamma>$ = 1.9) 
and the associated dispersion from the $\Gamma-$L$_{bol}$/L$_{Edd}$ relation \citep[$\sigma$ 
= 0.33, ][]{Risaliti09} are plotted as solid lines. }
\label{fig:fakeSED}
\end{figure}

\begin{figure}
\centering
\includegraphics[]{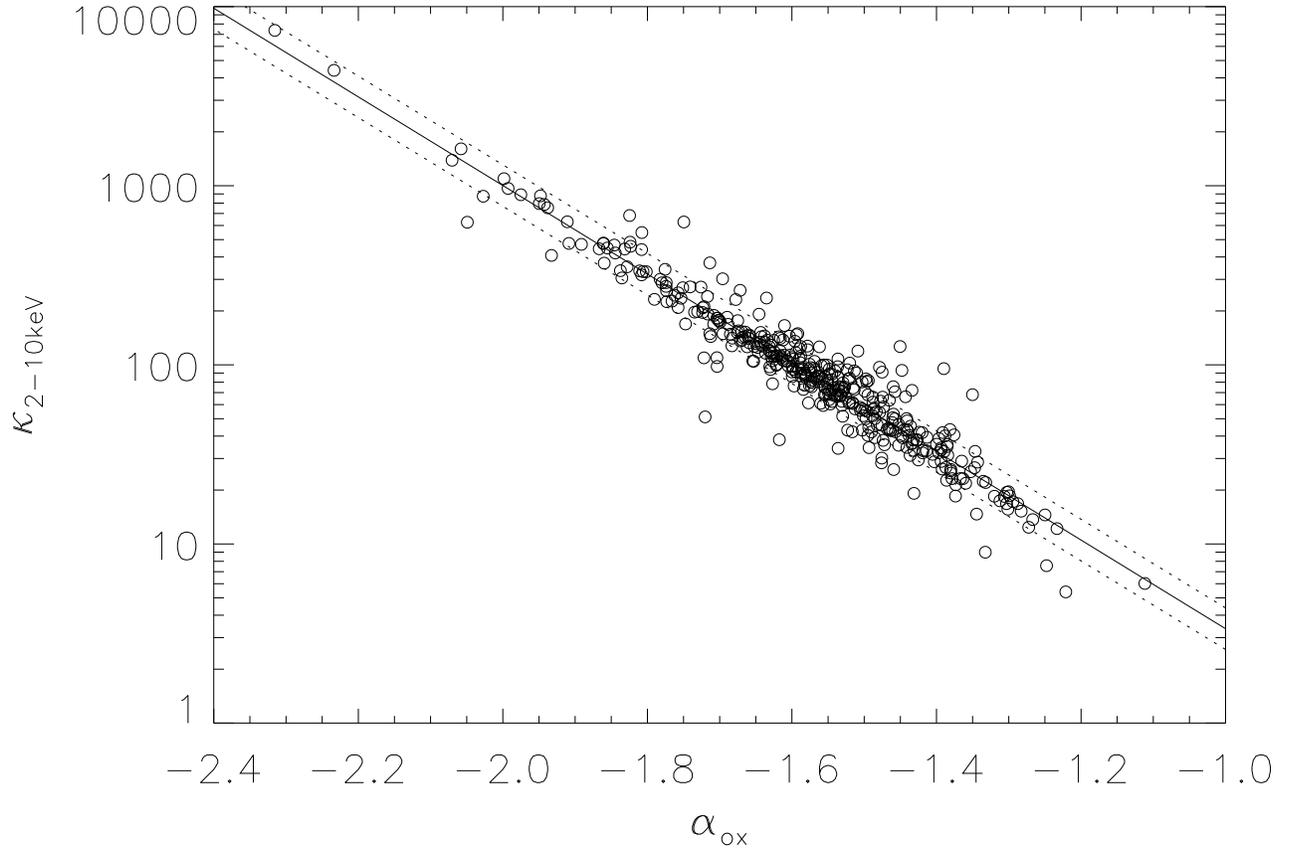}
\caption{X-ray bolometric correction vs. $\alpha_{ox}$.  The solid line is the best-fit, 
with the dispersion $\sigma(log \kappa)$ = 0.12 plotted as dotted lines.  Only detected 
(S/N $>$ 2), non-BAL, RQ sources with good X-ray fits are included.  Where X-ray S/N $>$ 6, 
sources with significant intrinsic absorption are excluded.  }
\label{fig:kbol}
\end{figure}

\begin{figure}
\centering
\includegraphics[]{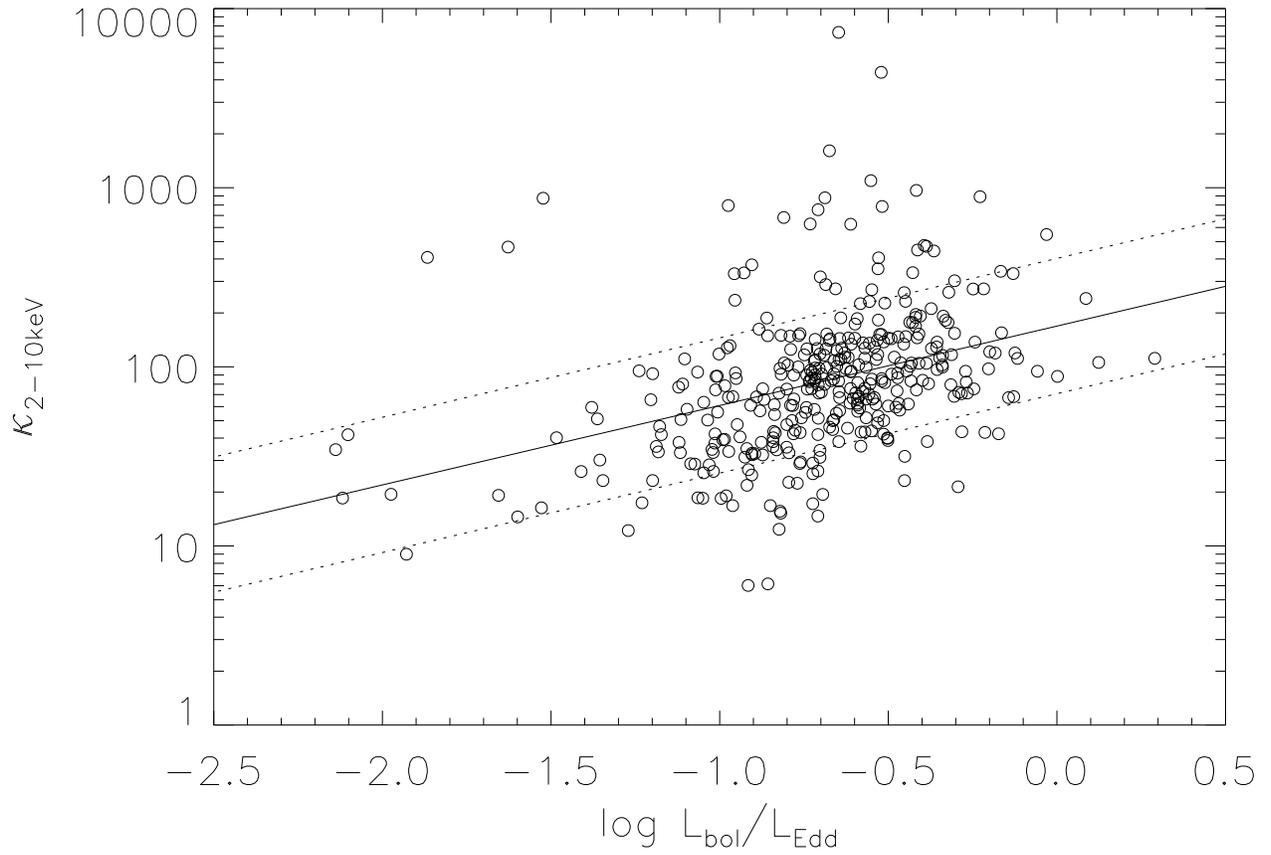}
\caption{X-ray bolometric correction vs. L$_{bol}$/L$_{Edd}$.  The solid line is the best-fit, 
with the dispersion $\sigma(log \kappa)$ = 0.38 plotted as dotted lines.  Sources included as 
in Figure \ref{fig:kbol}.}
\label{fig:kbolEdd}
\end{figure}


\begin{thebibliography}{}
\bibitem[Akritas \& Siebert(1996)]{AS96}  Akritas M.G. \& Siebert J.  1996, MNRAS, 278, 919
\bibitem[Anderson \& Margon(1987)]{AM87}  Anderson S.F. \& Margon B. 1987, \apj, 314, 111
\bibitem[Avni, Worrall \& Margon(1995)]{AWM95}  Avni Y., Worrall D.M. \& Margon W.A.  1995, \apj, 454, 673
\bibitem[Avni \& Tananbaum(1982)]{AT82}  Avni Y. \& Tananbaum H. 1982, \apj, 262, L17
\bibitem[Bechtold et al.(2003)]{Bechtold03} Bechtold J. et al.  2003, \apj, 588, 119
\bibitem[Beloborodov(1999)]{Beloborodov99}  Beloborodov A. M.  1999, \apj, 510, L123
\bibitem[Bevington et al.(1992)]{Bevington92}  Bevington, P. R. \& Robinson, D. K. 1992, Data Reduction 
and Error Analysis for the Physical Sciences (2d ed; New York: McGraw-Hill)
\bibitem[Boroson \& Green (1992)]{BG92}  Boroson, T. A., \& Green, R. F. 1992, ApJS, 80, 109
\bibitem[Brandt et al.(1997)]{Brandt97}  Brandt, W. N., Mathur, S., \& Elvis, M. 1997, MNRAS, 285, L25
\bibitem[Elvis et al.(1994)]{Elvis94}  Elvis M. et al.  1994, \apj, 95, 1 
\bibitem[Gibson et al.(2009)]{Gibson09}  Gibson et al.  2009, \apj, 692, 758
\bibitem[Green \& Mathur(1996)]{GM96}  Green P. \& Mathur S. 1996, \apj, 462, 637 
\bibitem[Haardt \& Maraschi(1991)]{HM91}  Haardt F. \& Maraschi L.  1991, \apj, 380, L51
\bibitem[Hopkins et al.(2007)]{Hopkins07}  Hopkins P.F., Richards G.T. \& Hernquist L.  2007, \apj, 654, 731
\bibitem[Isobe et al.(1986)]{Isobe86}  Isobe T., Feigelson E.D. \& Nelson P.I.  1986, \apj, 306, 490
\bibitem[Just et al.(2007)]{Just07}  Just D.W., Brandt W.N. Shemmer O., Steffen A.T., Schneider D.P., 
Chartas G. \& Garmire G.P.  2007, \apj, 665, 1004
\bibitem[Kawaguchi et al.(2001)]{Kawaguchi01}  Kawaguchi T., Shimura T. \& Mineshige S.  2001, \apj, 546, 966
\bibitem[Kelly et al.(2007)]{Kelly07}  Kelly B.C., Bechtold J., Siemiginowska A., Aldcroft T. \& Sobolewska M.  
2007, \apj, 657, 116
\bibitem[Kelly et al.(2007)]{Kelly08}  Kelly, B. C., Bechtold, J., Trump, J. R., Vestergaard, M., \& Siemiginowska, A. 
2008, ApJS, 176, 355
\bibitem[Kriss \& Canazares(1985)]{KC85} Kriss G.A. \& Canazares C.R. 1985, \apj, 297, 177
\bibitem[Lavalley et al.(1992)]{Lavalley92}  Lavalley M., Isobe T. \& Feigelson E.  1992, ASPC, 25, 245L
\bibitem[Mainieri et al.(2007)]{Mainieri07}  Mainieri et al. 2007, \apj, 172, 368
\bibitem[Malkan \& Sargent(1982)]{MS82}  Malkan M. \& Sargent W. 1982, \apj, 254, 22
\bibitem[Malzac(2001)]{Malzac01}  Malzac J. 2001, MNRAS, 325, 1625
\bibitem[Mateos et al.(2005)]{Mateos05}  Mateos et al. 2005, A\&A, 433, 855 
\bibitem[Nayakshin(2000)]{Nayakshin00}  Nayakshin S., Kazanas D. \& Kallman T.R.  2000, \apj, 537, 833
\bibitem[Pickering et al.(1994)]{PIF94} Pickering T.E., Imwanna pipey C.D. \& Foltz C.B. 1994, AJ, 108, 5
\bibitem[Prevot et al.(1984)]{Prevot84}  Prevot M., Lequeux J., Maurice E., Prevot L. \& 
Rocca-Volmerange B. 1984, A\&A, 132, 389
\bibitem[Proga(2007)]{Proga07}  Proga D. 2007, ApJ,661, 702
\bibitem[Richards et al.(2003)]{Richards03}  Richards et al. 2003, \aj, 126, 1131 
\bibitem[Richards et al.(2006)]{Richards06}  Richards et al. 2006, ApJS, 166, 470
\bibitem[Risaliti et al.(2009)]{Risaliti09}  Risaliti G., Young M. \& Elvis M.  2009, astroph/0906.1983
\bibitem[Shen et al.(2008)]{Shen08}  Shen Y., Greene J., Strauss M., Richards G. \& Schneider D.  2008, ApJ, 680, 169
\bibitem[Shemmer et al.(2006)]{Shemmer06}  Shemmer, O., Brandt, W. N., Netzer, H., Maiolino, R., \& Kaspi, S. 
2006, ApJ, 646, L29
\bibitem[Shemmer et al.(2008)]{Shemmer08}  Shemmer O., Brandt W.N., Netzer H., Maiolino R. \& Kaspi S. 
2008, ApJ, 682, 81
\bibitem[Shields et al.(1978)]{Shields78}  Shields G.A. 1978, Nature, 272, 706
\bibitem[Sobolewska et al.(2004a)]{Sobolewska04a}  Sobolewska M.A., Siemiginowska A., Zycki, P.T.  2004, ApJ, 608, 80
\bibitem[Sobolewska et al.(2004b)]{Sobolewska04b}  Sobolewska M.A., Siemiginowska A., Zycki, P.T.  2004, ApJ, 617, 102
\bibitem[Spergel et al.(2003)]{Spergel03}  Spergel D.N. et al. 2003, ApJS, 148, 175
\bibitem[Steffen et al.(2006)]{Steffen06}  Steffen A., Strateva I., Brandt W., Alexander D., Koekemoer A., 
Lehmer B., Schneider D. \& Vignali C. 2006, \aj, 131, 2826
\bibitem[Strateva et al.(2005)]{Strateva05}  Strateva I.V., Brandt W.N., Schneider D.P., Vanden Berk D.G. 
\& Vignali C. 2005, \apj, 130, 387
\bibitem[Tananbaum et al.(1986)]{Tananbaum86}  Tananbaum H., Avni Y., Green R.F., Schmidt M. \& Zamorani G. 
1986, \apj, 305, 57
\bibitem[Tang et al.(2007)]{Tang07}  Tang S.M., Zhang S.N. \& Hopkins P.  2007, MNRAS, 377, 1113
\bibitem[Vanden Berk et al.(2001)]{VdB01}  Vanden Berk D.E. et al. 2001, \aj, 122, 549
\bibitem[Vasudevan \& Fabian(2009)]{VF09}  Vasudevan, R.V. \& Fabian A.C.  2009, MNRAS, 392, 1124
\bibitem[Vasudevan et al.(2009)]{Vasudevan09}  Vasudevan, R.V., Mushotzky R.F., Winter L.M., \& Fabian A.C.  2009, MNRAS, 399, 1553
\bibitem[Vignali et al.(2003)]{Vignali03}  Vignali C., Brandt W.N. \& Schneider D.P. 2002, \aj, 125, 443 
\bibitem[Ward et al.(1987)]{Ward87}  Ward M.J., Elvis M., Fabbiano G., Carleton N.P., Willner S.P. \& Lawrence A.  
1987, \apj, 315, 74
\bibitem[Wilkes et al.(1994)]{Wilkes94}  Wilkes B.J., Tananbaum H., Worrall D.M., Avni Y., Oey M.S. \& Flanagan J.
1994, ApJS, 92, 52
\bibitem[Wills, Netzer \& Wills (1985)]{WNW85} Wills B.J., Netzer H. \& Wills D. 1985, /apj, 288, 94
\bibitem[Young et al.(2009)]{Young09}  Young M., Elvis M. \& Risaliti G.  2009, ApJS, 183, 17
\bibitem[Yuan et al.(1998)]{Yuan98}  Yuan W., Siebert J. \& Brinkmann, W.  1998, AA, 394, 498
\bibitem[Zdziarski et al.(1999)]{Zdziarski99}   Zdziarski A. A., Lubinski, P., \& Smith, D. A.  1999, MNRAS, 303, L11
\bibitem[Zdziarski et al.(2000)]{Zdziarski00}   Zdziarski A. A., Poutanen J. \& Johnson W.N.  2000, \apj, 542, 703
\end{thebibliography}
\end{document}